\documentclass[fleqn,usenatbib]{mnras}

\usepackage[T1]{fontenc}
\usepackage{ae,aecompl}

\usepackage{graphicx}	
\usepackage{amsmath}	
\usepackage{amssymb}	
\usepackage{longtable} 
\usepackage{hyperref}
\usepackage{cleveref}

\newcommand{\gaia}{{\it Gaia}}

\title[Gaia-IPHAS/KIS value-added catalogues]{The Gaia/IPHAS and Gaia/KIS Value-Added Catalogues}

\author[S. Scaringi et al.]{S. Scaringi$^{1,2}$\thanks{E-mail: simone.scaringi@ttu.edu}, 
C. Knigge$^{3}$,
J.~E. Drew$^{4}$,
M. Mongui\'o$^{4}$,
E. Breedt$^{5}$,
M. Fratta$^{1,2}$,
\newauthor
B. G{\"a}nsicke$^{6}$,
T.~J. Maccarone$^{1}$,
A.~F. Pala$^{6}$,
C. Schill$^{2}$
\\
$^{1}$Department of Physics and Astronomy, Texas Tech University, Lubbock, Texas 79409-1051, USA\\
$^{2}$School of Physical and Chemical Sciences, University of Canterbury, Christchurch 8041, New Zealand\\
$^{3}$School of Physics and Astronomy, University of Southampton, Highfield, Southampton SO17 1BJ, UK\\
$^{4}$School of Physics, Astronomy and Mathematics, University of Hertfordshire, College Lane, Hatfield, Hertfordshire AL10 9AB, UK\\
$^{5}$Institute of Astronomy, University of Cambridge, Madingley Road, Cambridge, CB3 0HA, United Kingdom\\
$^{6}$Astronomy and Astrophysics Group, Department of Physics, University of Warwick, Gibbet Hill Road, Coventry, CV4 7AL, UK\\
}

\date{Accepted XXX. Received YYY; in original form ZZZ}

\pubyear{2018}

\begin{document}
\label{firstpage}
\pagerange{\pageref{firstpage}--\pageref{lastpage}}
\maketitle

\begin{abstract}
We present a sub-arcsecond cross-match of \gaia\ DR2 against the INT Photometric H$\alpha$ Survey of the Northern Galactic Plane Data Release 2 (IPHAS DR2) and the \textit{Kepler}-INT Survey (KIS). The resulting value-added catalogues (VACs) provide additional precise photometry to the \gaia\ photometry ($r$, $i$ and H$\alpha$ for IPHAS, with additional $U$ and $g$ for KIS).  In building the catalogue, proper motions given in \gaia\ DR2 are wound back to match the epochs of IPHAS DR2, thus ensuring high proper motion objects are appropriately cross-matched. The catalogues contain 7,927,224 and 791,071 sources for IPHAS and KIS, respectively. The requirement of $>5\sigma$ parallax detection for every included source means that distances out to 1--1.5 kpc are well covered. We define two additional parameters for each catalogued object: (i) $f_c$, a magnitude-dependent tracer of the quality of the \gaia\ astrometric fit; (ii) $f_{FP}$, the false-positive rate for parallax measurements determined from astrometric fits of a given quality at a given magnitude. Selection cuts based on these parameters can be used to clean colour-magnitude and colour-colour diagrams in a controlled and justified manner. We provide both full and light versions of the VAC, with VAC-light containing only objects that represent our recommended trade-off between purity and completeness.  Uses of the catalogues include the identification of new variable stars in the matched data sets, and more complete identification of H$\alpha$-excess emission objects thanks to separation of high-luminosity stars from the main sequence.
\end{abstract}




\section{Introduction}
\begin{figure*}
	\includegraphics[width=2\columnwidth]{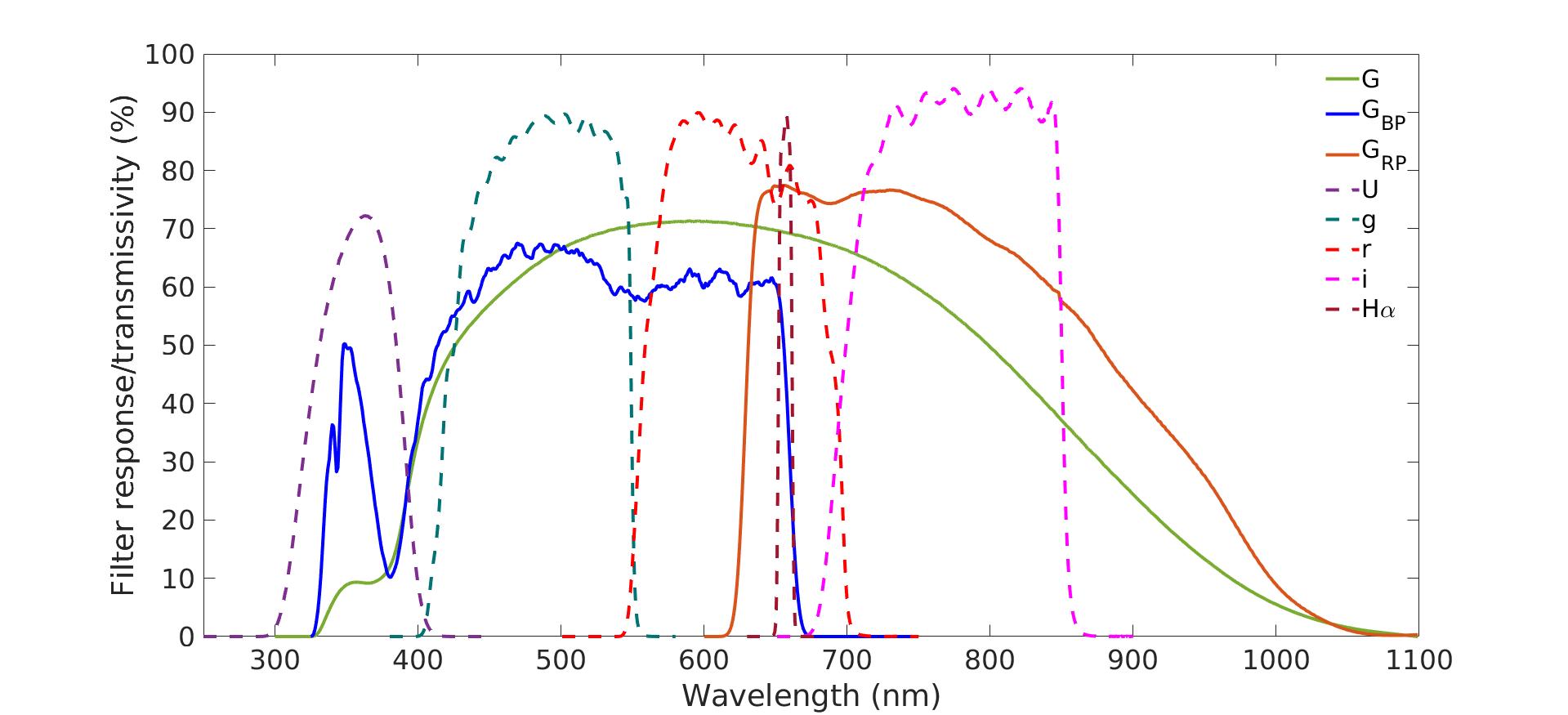}
    \caption{Total response curves from \gaia\ ($G$, $G_{BP}$, $G_{RP}$), and filter transmission curves from IPHAS ($r$, $i$, H$\alpha$) and KIS ($U$, $g$, $r$, $i$, H$\alpha$). }
    \label{fig:bands}
\end{figure*}

The European Space Agency's \gaia\ mission provides unprecedented opportunities to assemble reliable Hertzsprung-Russell diagram for different types of stellar populations. \gaia\ Data Release 2 (DR2; \citealt{gaia,gaiaDR2,gaiaCMD,gaiaAstro,arenou18}) includes photometry in the $G$, $G_{BP}$ and $G_{RP}$ bands (see Figure \ref{fig:bands}) for approximately 1.5 billion sources. Its quality and size will define the new standard in the years to come, and have a tremendous impact on various areas of astrophysics. In particular, it is the astrometry, and specifically the parallax measurements, that will provide the largest impact, since it is with these measurements that we can now infer distances, absolute magnitudes, and transverse velocities (with the additional proper motion information) for individual targets.

The INT/WFC Photometric H$\alpha$ Survey of the Northern Galactic Plane (IPHAS; \citealt{drew05}) is the first comprehensive digital survey of the northern Galactic disc ($|b|<5^\circ$), covering a Galactic longitude range of $29^\circ < l < 215^\circ$. The IPHAS observations are obtained using the Wide Field Camera (WFC) at the prime focus of the 2.5m Isaac Newton Telescope (INT) on La Palma, Spain. IPHAS images are taken through three filters: a narrow-band H$\alpha$, and two broad-band Sloan $r$ and $i$ filters (see Figure \ref{fig:bands}). Exposures are set to reach an $r$-band depth of $\approx 21$. Pipeline data reduction is handled by the Cambridge Astronomical Survey Unit (CASU). Further details on the data acquisition and pipeline reduction can be found in \cite{drew05} and \cite{gonzalez08}. In this paper we use the IPHAS Data Release 2 (\citealt{IPHASDR2}), containing measurements for $\approx 219$ million sources observed between 2003-2012 which have all been photometrically calibrated.

The \textit{Kepler}-INT Survey (KIS; \citealt{KIS}) observed the \textit{Kepler} field using the same observing strategy as IPHAS on the INT telescope, with additional observations in the Sloan $g'$ band and non-standard U-band (see Figure \ref{fig:bands}), both of which are also used in the UV-Excess Survey of the Northern Galactic Plane (UVEX; \citealt{uvex}). The pipeline data reduction is identical to IPHAS. In this paper we use the KIS Data Release 2, which provides coverage of 97\% of the \textit{Kepler} field, and contains $\approx14.5$ million photometrically calibrated entries. 

This paper presents sub-arcsecond cross-matches between the \gaia/IPHAS and \gaia/KIS catalogues, taking into account the different epochs of observations of both IPHAS and KIS, as well as the proper motion information for each target in the \gaia\ catalogue. In the process of producing these catalogues we additionally include, for each target, two additional columns: a so-called ``completeness'' fraction ($f_c$) which provides information relating to how acceptable the \gaia\ astrometric solution is compared to targets with similar $G$-band magnitudes, and a so-called false-positive fraction ($f_{FP}$) providing information on how reliable the astrometric measurements of a given target are.

Section \ref{sec:crossMatch} describes our cross-matching procedure, including the preliminary selection cuts applied to all datasets and examples of recovered matches. Section \ref{sec:addPars} introduces our additional quality control parameters $f_c$ and $f_{FP}$, and discusses how these can be used to clean the \gaia/IPHAS and \gaia/KIS catalogues from unreliable entries. Section \ref{sec:science} provides some illustrative examples of how our value-added catalogues can be used for science exploitation. Finally Section \ref{sec:VAC} describes our published catalogue formats, with conclusions drawn in Section \ref{sec:conc}.

\section{Cross-matching Gaia with IPHAS and KIS} \label{sec:crossMatch}
The \gaia\ DR2 release contains results for over 1.6 billion sources. The majority of this data is not required for our cross-matching purposes since it either lies outside the IPHAS/KIS footprint and/or the \gaia\ results are not of high enough quality. In this section we describe how we performed a sub-arcsecond matching between the IPHAS/KIS targets with \gaia\ DR2, including descriptions of the selection cuts and proper motion corrections. We also highlight the advantage of our method against a simple cross-match through some examples.   

\subsection{Selection cuts} \label{sec:selection}
Before attempting to cross-match sources in IPHAS/KIS with \gaia\ DR2, we apply some quality cuts to all datasets in order to retain only sources with good photometric and astrometric measurements.

From IPHAS DR2 we select only objects which:
\begin{itemize}
\item have measurements in all three bands ($r$, $i$ and H$\alpha$);
\item are fainter than the saturation limit in all 3 photometric bands ($r>13$, $i>12$ and H$\alpha>12.5$);
\item have photometric errors smaller or equal to 0.1 mag in all bands;
\item are not flagged as blended or affected by bright neighbours in any band.
\end{itemize}
\noindent Of the 218,991,524 sources in IPHAS DR2 63,520,381 survive these quality cuts. Similar cuts are applied to the KIS DR2 catalogue, with the inclusion of the same cuts in the $U$ and $g$ bands. This retained 2,662,117 sources out of 14,476,957.

From \gaia\ DR2 we select only objects which:
\begin{itemize}
\item have a $G$-band flux signal-to-noise above 5 (\texttt{phot\_g\_mean\_flux\_over\_error}>5);
\item have a signal-to-noise parallax measurement above 5 (\texttt{parallax\_over\_error}>5);
\item are within an area slightly larger than the IPHAS footprint ($20<l<220$ and $-6<b<6$);
\item are within the KIS footprint ($275<\alpha_{J2015.5}$<305 and $36<\delta_{J2015.5}<54$).
\end{itemize}
\noindent The above selection criteria yield two \gaia\ datasets: one containing 19,553,253 sources within the IPHAS footprint, and one containing 3,004,331 sources in the KIS footprint.

\subsection{Proper motion corrections and cross-matching} \label{sec:matching}
In order to minimise mismatches between the \gaia\ catalogue and both IPHAS and KIS, as well as recovering fast moving objects, it is important to take into account the proper motion of targets. \gaia\ DR2 provides proper motion information for all targets within the IPHAS and KIS footprints which pass our data quality cuts. However, given the way they were designed, neither IPHAS and KIS contain this information. Furthermore, although all catalogues give positions in the barycentric ICRS reference frame, only the \gaia\ DR2 positions are given at epoch 2015.5. Both IPHAS and KIS report positions at the epoch of observation, which can be any time between 2003 -- 2012 for IPHAS and 2011 -- 2012 for KIS. The epoch of observation for a particular target is reported in both IPHAS and KIS DR2 catalogues.

Ideally, for precise cross-matching between the catalogues, the \gaia\ astrometry would have to be propagated to the IPHAS/KIS epoch of each source before the cross-matching is performed. This would result in recomputing the \gaia\ astrometry for every entry in the input catalogues (in excess of 20 million when considering both IPHAS and KIS), and becomes even more unfeasible for larger input catalogues. Instead we proceed by first dividing the IPHAS and KIS cleaned catalogues into monthly epoch batches. Because of the observing strategy of both IPHAS and KIS, which sequentially observe all bands immediately following each other, we take the epoch of a particular target to be the start of the $r$-band observation. This ensures that the epoch-corrected positional uncertainty of the \gaia\ catalogue is relativity small even for high proper motion objects. For example, the recomputed \gaia\ coordinates for an object with an extreme proper motion of 2''/year will be at worst $\approx0.08$'' off the IPHAS/KIS position.

This procedure results in 46 monthly batches for IPHAS and 6 for KIS. For each of these batches, we then recompute the \gaia\ astrometry to the mid-point epoch for each month. We then select the positional closest match in the sky within 1'' of a given IPHAS or KIS entry. After removing for \gaia\ duplicated sources\footnote{Targets flagged as duplicate sources in the \gaia\ archive may indicate observational, cross-matching or processing problems, or stellar multiplicity, and probable astrometric or photometric problems in all cases.}, this retains 7,927,532 and 827,989 sources for the IPHAS and KIS footprints, respectively. However, because some areas of the sky in both IPHAS and KIS have been observed more than once (excluding offset fields) some of the retained sources will have duplicated entries in our catalogues. We thus clean the retained sources by removing duplicates based on their \gaia\ DR2 designation. The retained number of sources is then 7,927,224 and 791,071 for IPHAS and KIS respectively. 

To ensure that the correct match is found in cases where 2 or more targets are within the 1'' cross-match radius, we retained all matches found within 1'' when cross-matching IPHAS to \gaia. In total there are 3,253 pairs (no triples or more) which can be found within 1'' when doing the cross-match. We chose to then inspect the $G_{RP}-i$ colour for these to determine whether this information could help reduce any false-positive matches. Although the \gaia\ $G_{RP}$ fluxes are derived from simple integration of 3.5 by 2.1 arcsec windows (and thus cannot resolve ambiguous matches), the $G_{RP}$ and $i$ combination has been chosen since the $i$-band is the only IPHAS band to fully reside within a \gaia\ band (see Figure \ref{fig:bands}). Of the 3,253 pairs, only 4 targets have an absolute $G_{RP}-i$ value $>1$ magnitude. In all 4 cases the correct match was identified as being the closer target. Visual inspection of the remaining targets also reveals the closest match to be the correct one. 

\subsection{Recovered matches and removed false-positives}
In order to investigate the efficiency of correcting for proper motion in epochs of monthly batches, we have cross-matched the cleaned IPHAS and KIS catalogues to the cleaned \gaia\ catalogue at epoch J2000, selecting the closest match within a generous 5'' radius for every input target. This exercise mimics what would happen if one used the CDS XMatch service\footnote{\url{http://cdsxmatch.u-strasbg.fr/xmatch}} for cross-matching IPHAS DR2 and \gaia\ DR2. By comparing the results of this ``raw'' cross-match to the catalogue produced by our proper-motion corrected cross-match, we can identify false positives, false negatives and mismatches in the ``raw'' catalogues.

\begin{figure}
	\includegraphics[width=1\columnwidth]{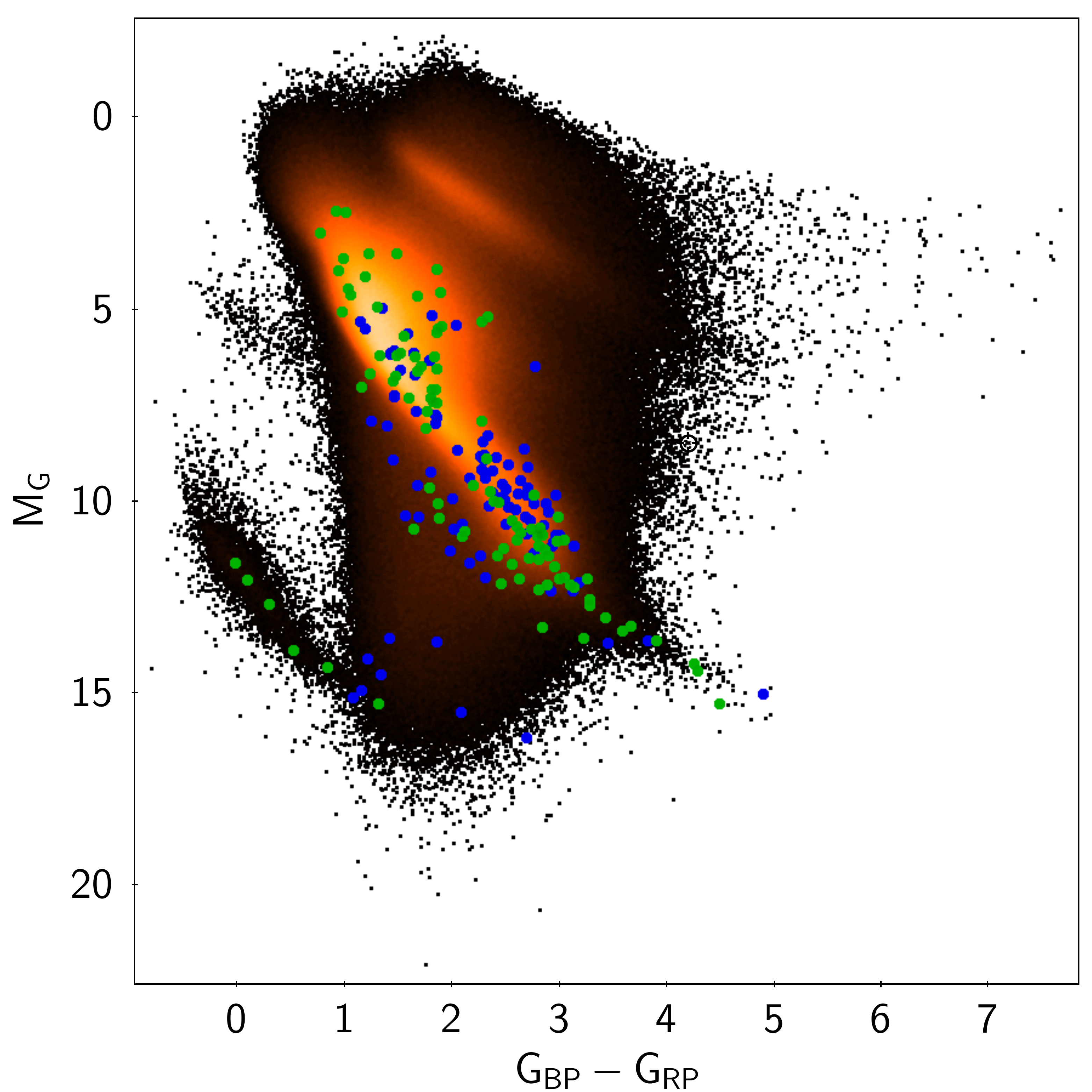}
    \caption{CMD from our cross-matched \gaia/IPHAS catalogue using \gaia\ photometry, adopting distances inferred from the \gaia\ parallaxes. The figure also displays objects which are only correctly recovered through the additional \gaia\ proper motion information. Objects which would have been entirely missed are marked in blue, whilst miss-matches are marked in green.}
    \label{fig:GaiaIPHAS_CMD}
\end{figure}

Figure \ref{fig:GaiaIPHAS_CMD} shows some colour-magnitude diagrams (CMD) for the retained sources within the IPHAS footprint. We note that all distances inferred in this paper have been determined via $M = m+5+5 \log_{10}(\varpi/1000)$, where $M$ and $m$ are the absolute and apparent magnitudes respectively, and $\varpi$ the parallax in milliarcseconds (the same practice as presented by \citealt{gaiaCMD}). These have been computed to generate the CMDs, and are not used for any selection or cross-matching purposes. We have also computed distances using the Exponentially Decreasing Space Density (EDSD) approach (\citealt{bj18,luri18}), adopting a scale height of $L=1.35$ kpc, but the results are qualitatively similar. We point out that $M$ and colour for all objects in Figure \ref{fig:GaiaIPHAS_CMD}, as well as all other CMDs plotted in this paper, have not been corrected for extinction, and hence are upper limits on true absolute magnitude and colour.

\begin{figure*}
	\includegraphics[width=1\columnwidth]{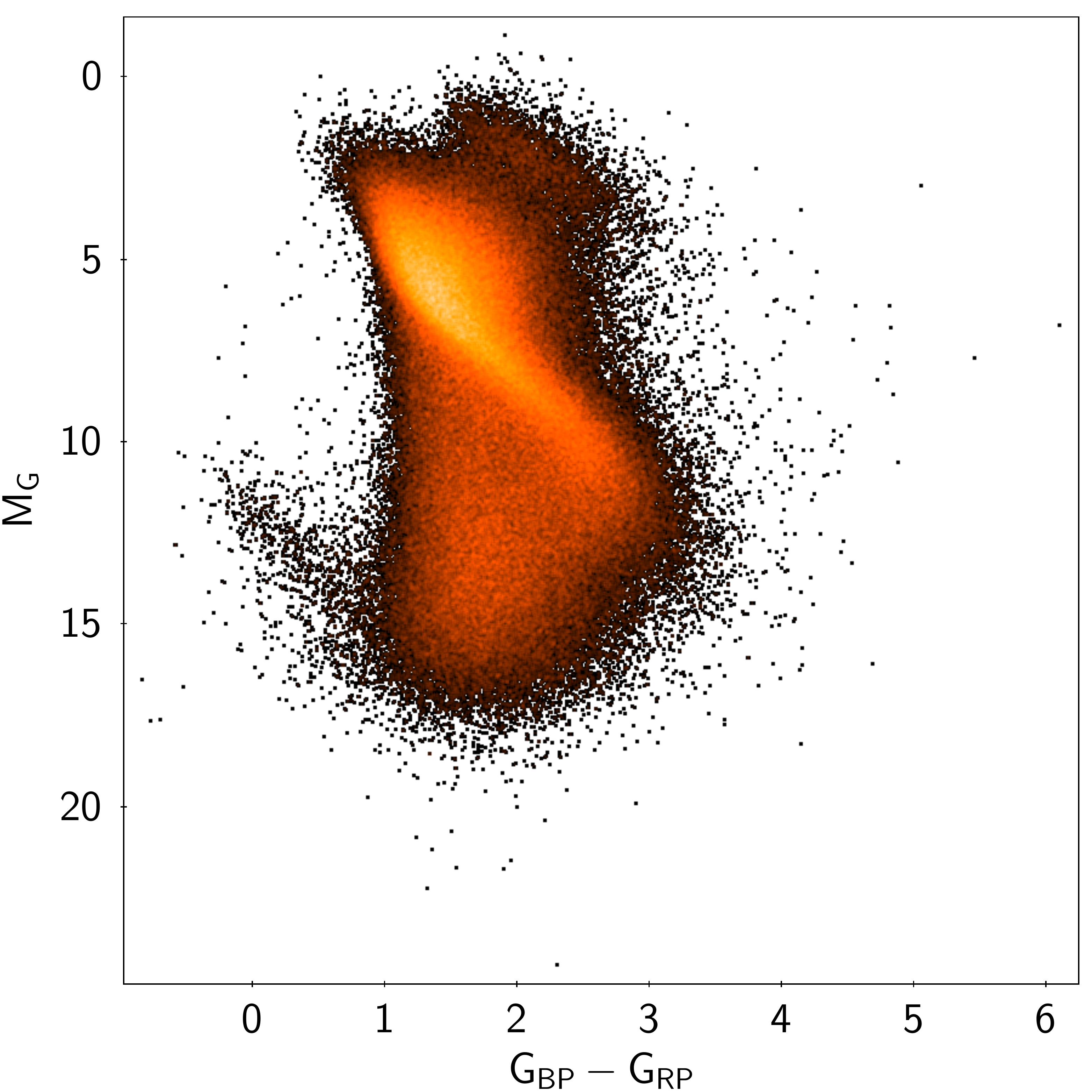}
    \includegraphics[width=1\columnwidth]{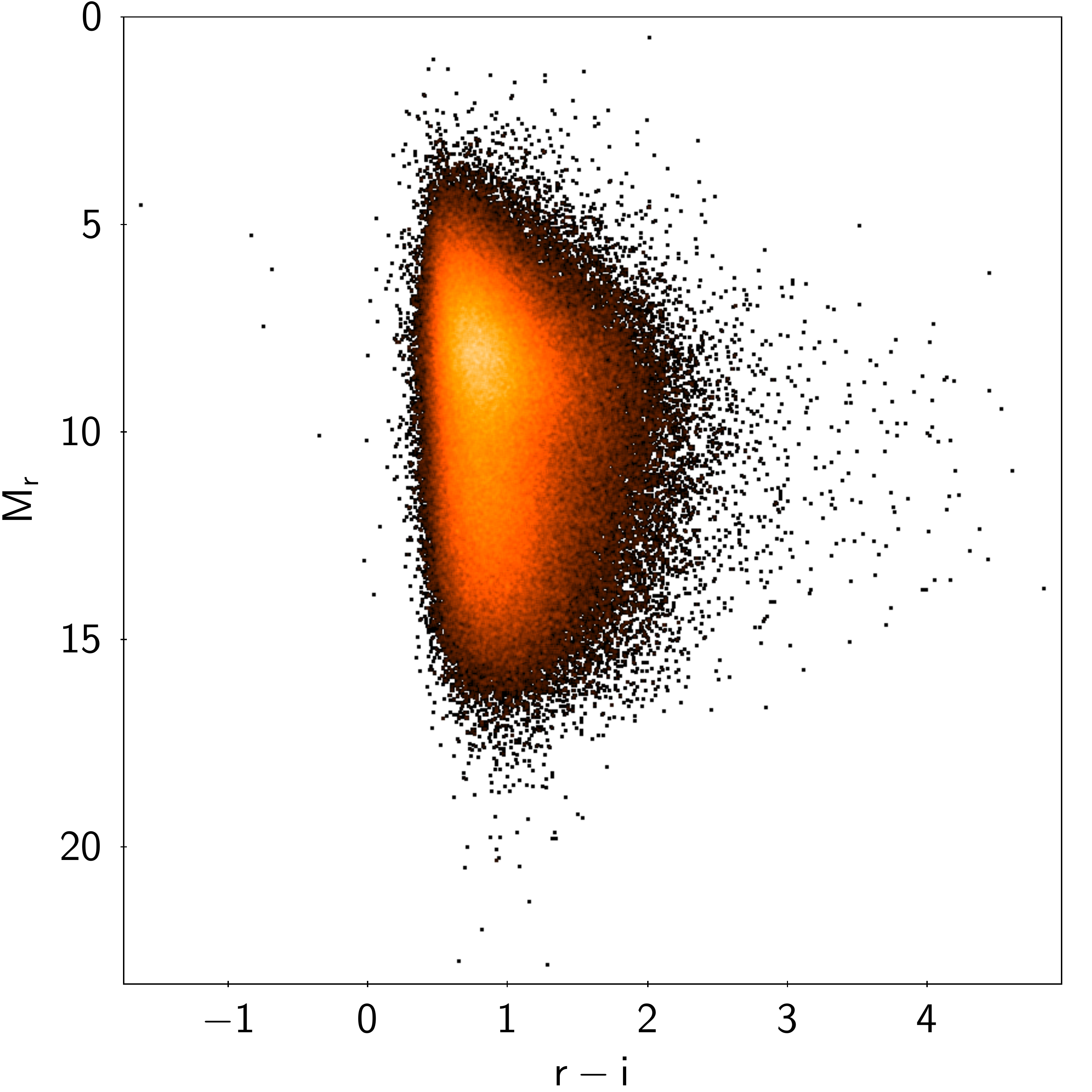}
    \caption{CMDs for IPHAS objects which would have been wrongly been associated to a \gaia\ counterpart using a standard 5'' cross-match radius, and ignoring proper motion information. The left panel displays the \gaia\ CMD, whilst the right panel displays the corresponding IPHAS CMD for the same sources.}
    \label{fig:GaiaIPHAS_CMD_BAD}
\end{figure*}

We recover 103 sources which would have been entirely missed by employing a simple 5'' search radius without proper motion correction (blue circles), whilst 101 sources would have been miss-matched (green circles). More importantly, 209,307 false-positive sources would have erroneously been included. The location of these mismatched sources in the CMD are shown in Figure \ref{fig:GaiaIPHAS_CMD_BAD}. Although the \gaia-based CMD shows some targets on the main sequence, the IPHAS-based CMD clearly shows that these false-positives do not lie on any known sequence (due to the erroneous match between the IPHAS colours and the \gaia\ distances), but rather occupy a region in parameter space which is known to be populated by stars with problematic distance estimates (\citealt{gaiaAstro}). 

\begin{figure*}
	\includegraphics[width=1\columnwidth]{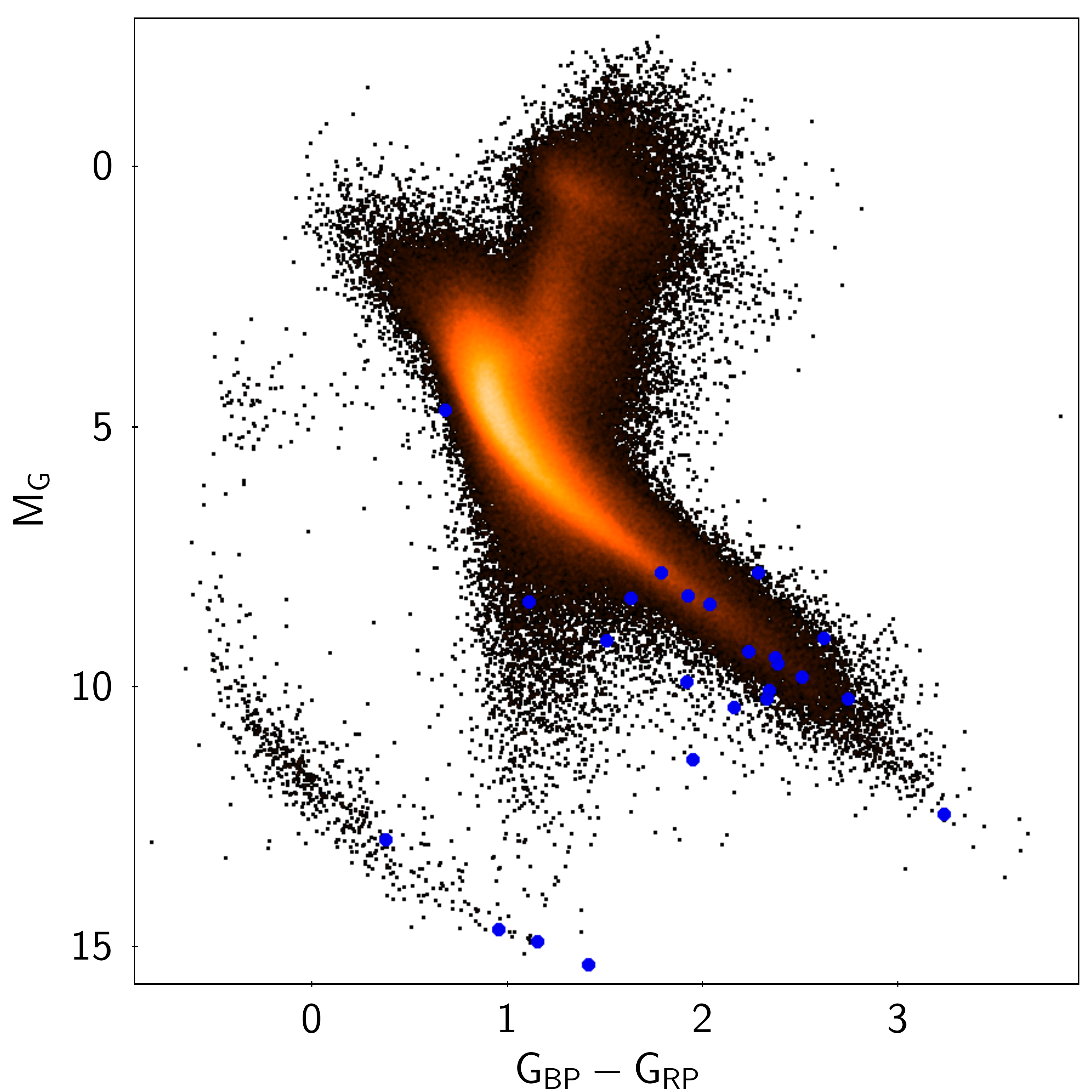}
    \includegraphics[width=1\columnwidth]{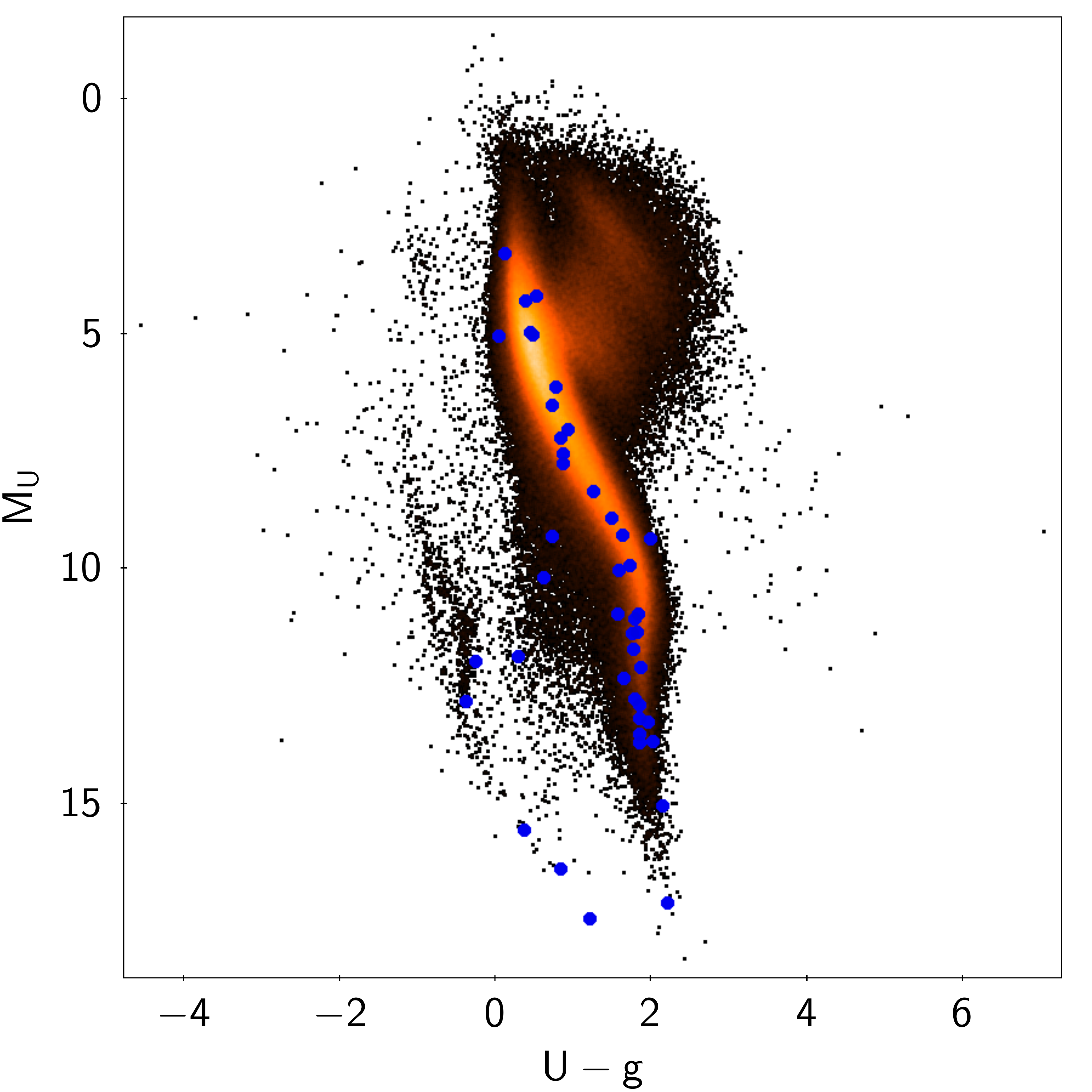}
    \caption{CMDs from our cross-matched \gaia/KIS catalogue using \gaia\ photometry (left panel) and KIS photometry ($U$ and $g$, right panel), both adopting distances inferred from the \gaia\ parallaxes. Both panels also display objects which would have been entirely missed without correcting for proper motion.}
    \label{fig:GaiaKIS_CMD}
\end{figure*}

Figure \ref{fig:GaiaKIS_CMD} shows some CMDs for our matched targets in the KIS footprint, including 43 targets which would have been missed without taking proper motion to account (blue circles). It is worth noting that the inclusion of the $U$ and $g$ bands from KIS clearly separates the H and He white dwarf tracks, as has been shown with SDSS colours in \cite{gaiaCMD}. We have omitted displaying 35,639 targets which have been miss-matched for clarity, but these show qualitatively similar problems as those shown in Figure \ref{fig:GaiaIPHAS_CMD_BAD} from the IPHAS sample. 

\begin{figure}
	\includegraphics[width=1\columnwidth]{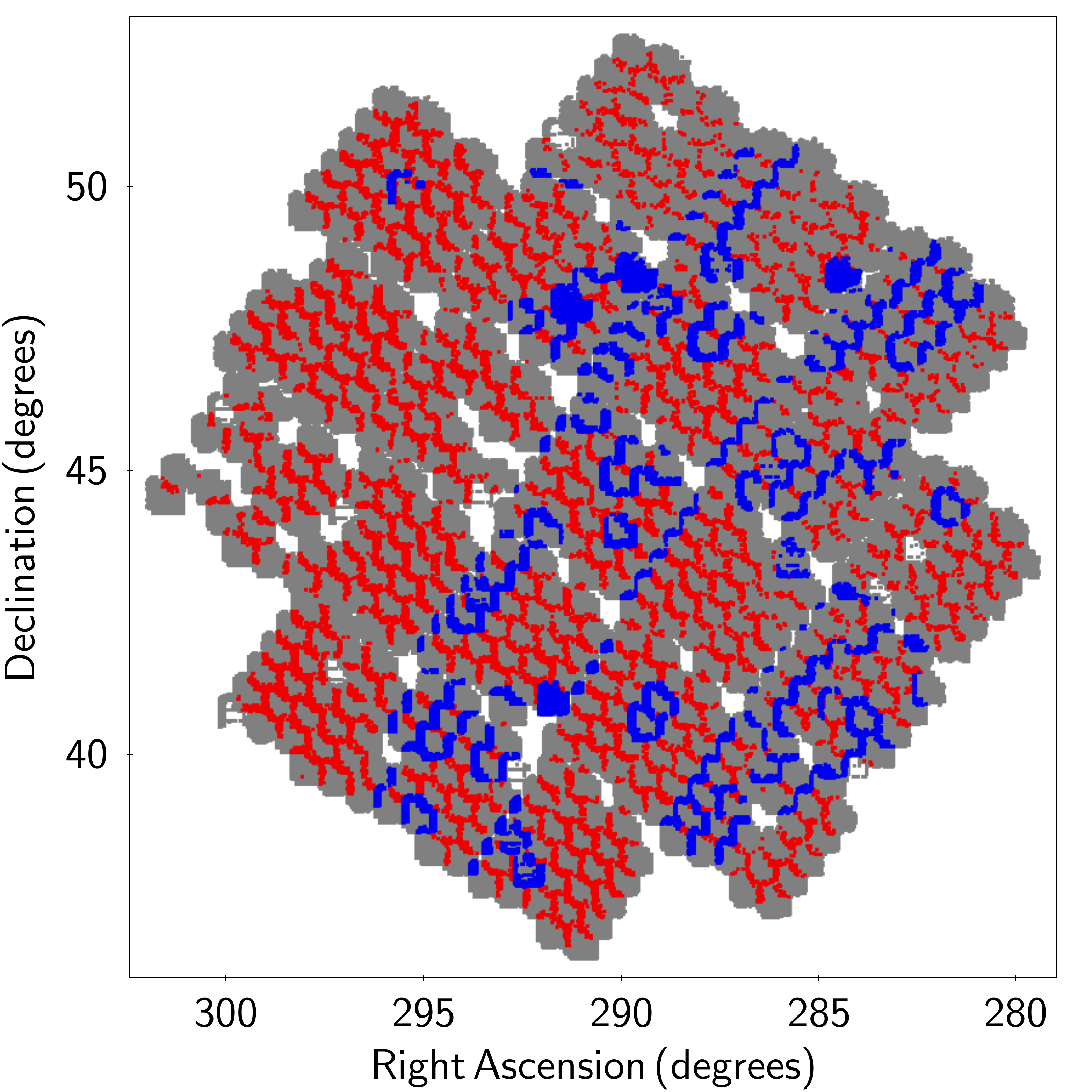}
	\caption{Sky position for all targets retained in our \gaia/KIS catalogue (grey points). Overlayed are targets which would have been miss-matched using a standard 5'' cross-match radius (red points), as well as targets which appeared as duplicates in our final \gaia/KIS catalogue due to being observed multiple times during the KIS survey (blue points).}
    \label{fig:KIS_footprint}
\end{figure}

It is interesting that the number of mismatches, as well as the number of duplicates (36,184), is relatively higher in the uncorrected \gaia/KIS catalogue than in the uncorrected \gaia/IPHAS one. This is due to the way the \textit{Kepler} footprint was tiled in KIS, with some areas being observed multiple times. This is illustrated in Figure \ref{fig:KIS_footprint}, which shows the sky position of the \gaia/KIS catalogue with the mismatches and duplicate sources displayed in red and blue respectively. The outline of the overlap regions is clearly visible. 

\begin{figure*}
\end{figure*}

\subsection{Bias induced by selection cuts}
The selection cuts described in Section \ref{sec:selection} will introduce a number of mismatches between the IPHAS/KIS catalogues and the epoch-corrected \gaia\ catalogue. In particular, the selection on \texttt{phot\_g\_mean\_flux\_over\_error}>5 will have a large effect, since it leads to many real \gaia\ detections not being available for cross-matching. As a result, a given IPHAS/KIS target may be matched to the wrong \gaia\ counterpart, as the true counterpart might have not passed our initial cuts. To estimate how large this effect is in our catalogues we re-ran our cross-matching algorithm on the full \gaia\ DR2 catalogue, without imposing any selection cuts. In order to keep this task manageable, we performed this exercise only in a moderately dense region of the IPHAS footprint (\citealt{farnhill16}), with $60<l<70$. Estimating the mismatch fraction in this region will then yield an upper limit on the mismatch fraction throughout both the \gaia/IPHAS and \gaia/KIS catalogues.

In total, there are over 27 million \gaia\ targets within $60<l<70$ and $-6<b<6$. Of these, $\approx 6$ million entries can be matched to an IPHAS source within 1''. Our \gaia/IPHAS catalogue contains 624,117 within the same footprint area, and we find 726 sources to have been mismatched based on their \gaia\ DR2 designation. We can thus place an upper limit of 0.1\% on the fraction of mismatches associated with our selection cuts, and note that many of these mismatches are very faint ($G>19$) \gaia\ sources, and that this effect will be much lower for the \gaia/KIS catalogue given the lower crowding of sources above the Galactic plane.

\section{Purity vs. Completeness: Introducing additional quality parameters} \label{sec:addPars}
In this section we will introduce some additional quality parameters which can be used to clean our merged \gaia/IPHAS and \gaia/KIS catalogues from targets with unreliable parallax measurements.

Even though we will only explicitly discuss the \gaia/IPHAS cross-match below, the same procedure has also been applied to the \gaia/KIS catalogue, with similar results.

\subsection{Completeness} \label{sec:completeness}
Ideally, sources from the \gaia\ catalogue that have poor astrometric solutions can be removed using the goodness-of-fit statistic provided by the \gaia\ Archive \citep[see discussion in ][]{gaiaAstro}. Rather than remove targets, we have opted to retain all sources, and instead include for each source a ``completeness'' value representing how good/bad the astrometric fit of a particular target is compared to targets within a similar apparent magnitude range (We will explain the reason for this choice of terminology below). 

\begin{figure}
	\includegraphics[width=1\columnwidth]{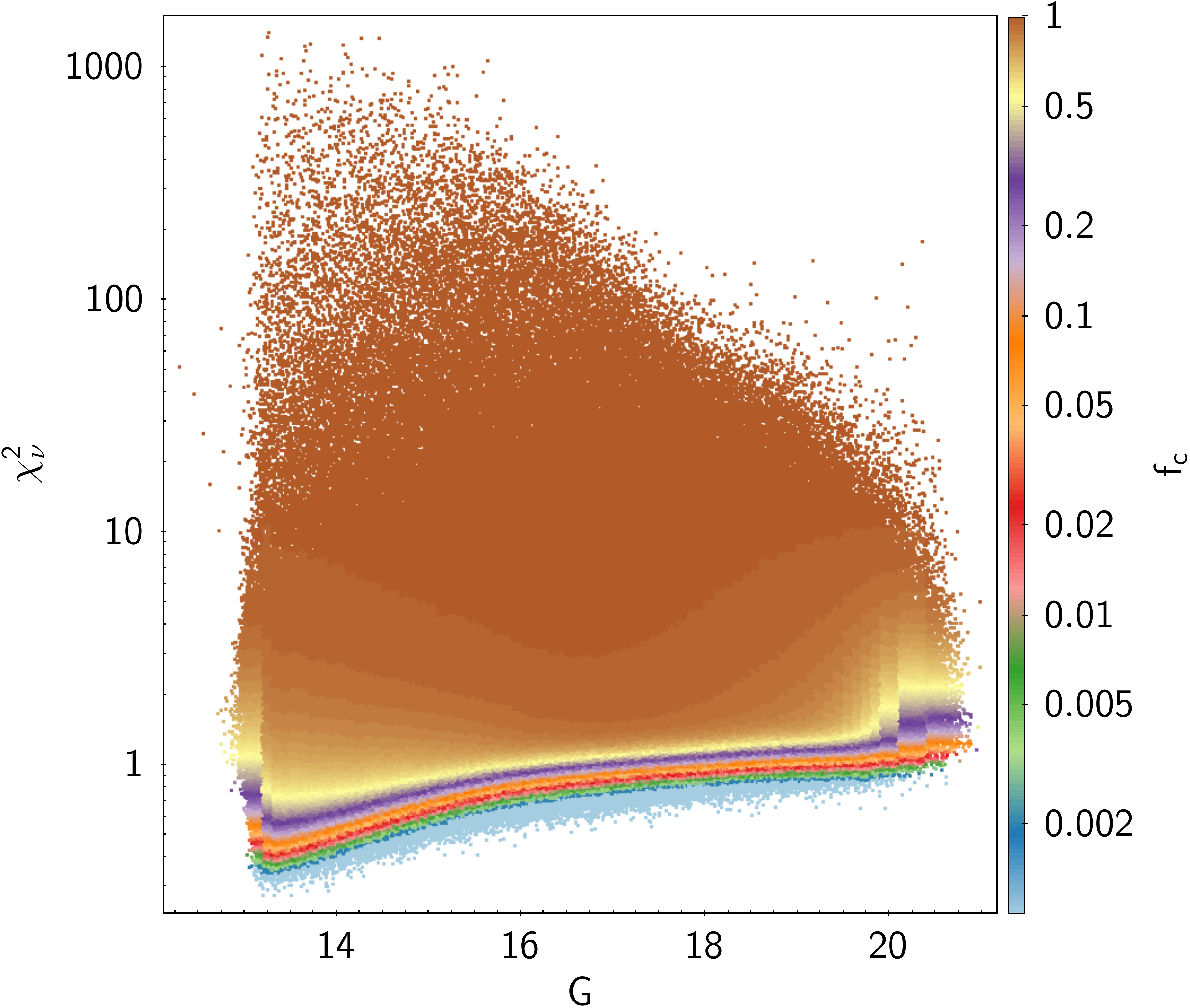}
    \caption{\gaia\ $G$-band magnitude versus reduced $\chi^2$ ($\chi_{\nu}^2$, based on the \gaia\ astrometric fit) for all targets retained in our \gaia/IPHAS catalogue. Targets are colour coded by their ``completeness'' fraction $f_c$ (see Section \ref{sec:completeness} for details).}
    \label{fig:Gchi2_fc}
\end{figure}

To do this we first re-computed, for each target, the reduced $\chi^2$ as $\chi_{\nu}^2=\texttt{astrometric\_chi2\_al}/(\texttt{astrometric\_n\_good\_obs\_al}-5)$. We then binned all sources in increasing $G$-band bins of 0.1 magnitudes, with the requirement that each bin contains at least 10,000 sources. Each source was then assigned a percentile based on its $\chi_{\nu}^2$ within their corresponding $G$-band magnitude bin. We refer to this percentile as the ``completeness fraction'', $f_c$, because it permits the targeted removal of sources with poor astrometry while retaining any desired completeness fraction. For example, removing all sources with $f_c > 0.9$ removes exactly 10\% of all sources (for a completeness of 90\%). Figure \ref{fig:Gchi2_fc} shows the resulting ($G$,$\chi_{\nu}^2$) plane for targets in our \gaia/IPHAS catalogue colour-coded with $f_c$. The apparent hard edges are the result of our binning scheme, and could be improved with larger number of objects. 

\subsection{Purity}
It is known and well-documented (\citealt{gaiaAstro}), that \gaia\ DR2 contains some spurious measurements, most notably very large or negative parallaxes. These spurious results are usually traced back to internal cross-matching issues or resolved (or partially resolved) binaries, and are expected to be corrected for in future \gaia\ releases (\citealt{gaiaAstro}). To mitigate this, \cite{gaiaAstro} have inspected where targets with negative parallaxes fall in $G$-magnitude vs. $u$ parameter space, where $u = \sqrt{\chi_{\nu}^2}$.  They then used this to define a simple threshold cut in ($G$,$u$) space that is designed to preferentially remove spurious measurements.

Following a similar procedure to \cite{gaiaAstro} we have produced a ``mirror sample'' of our cross-match catalogue that includes {\em only} sources with spurious \gaia\ measurements that are nevertheless as ``convincing'' as those in our actual catalogue. In order to construct this mirror sample, we query the \gaia\ DR2 archive with the same criteria described in Section \ref{sec:selection}, except we change the \texttt{parallax\_over\_error}>5 condition to \texttt{parallax\_over\_error}<-5. This query resulted in 603,742 sources for IPHAS and 25,295 sources for KIS. We then parsed our mirror sample through the same cross-matching procedure described in Section \ref{sec:matching}. This allows us to have a sample for which we know the astrometry is bad, but still retains the statistical properties of our catalogues. 

\begin{figure*}
	\includegraphics[width=1\columnwidth]{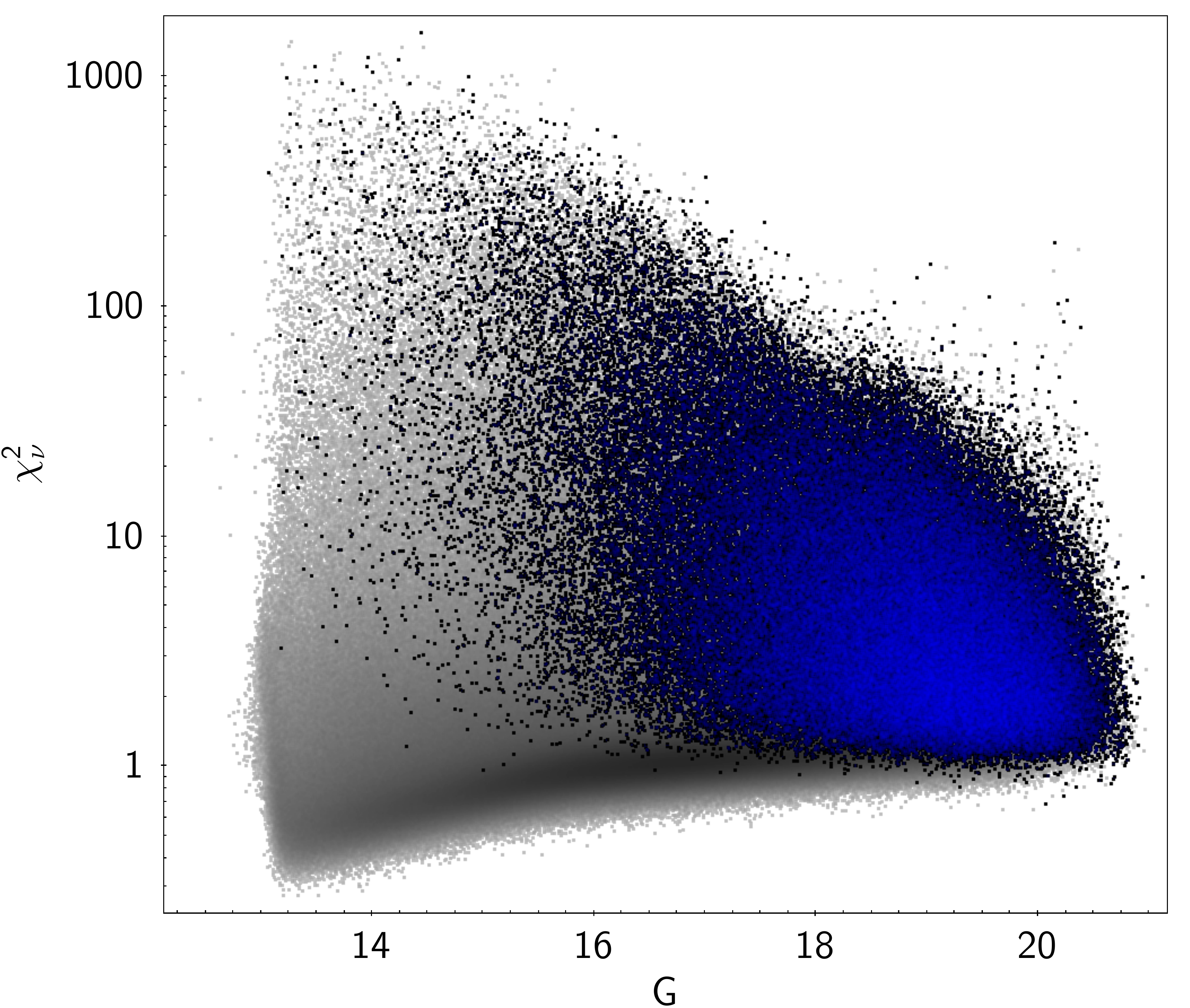}
    \includegraphics[width=1\columnwidth]{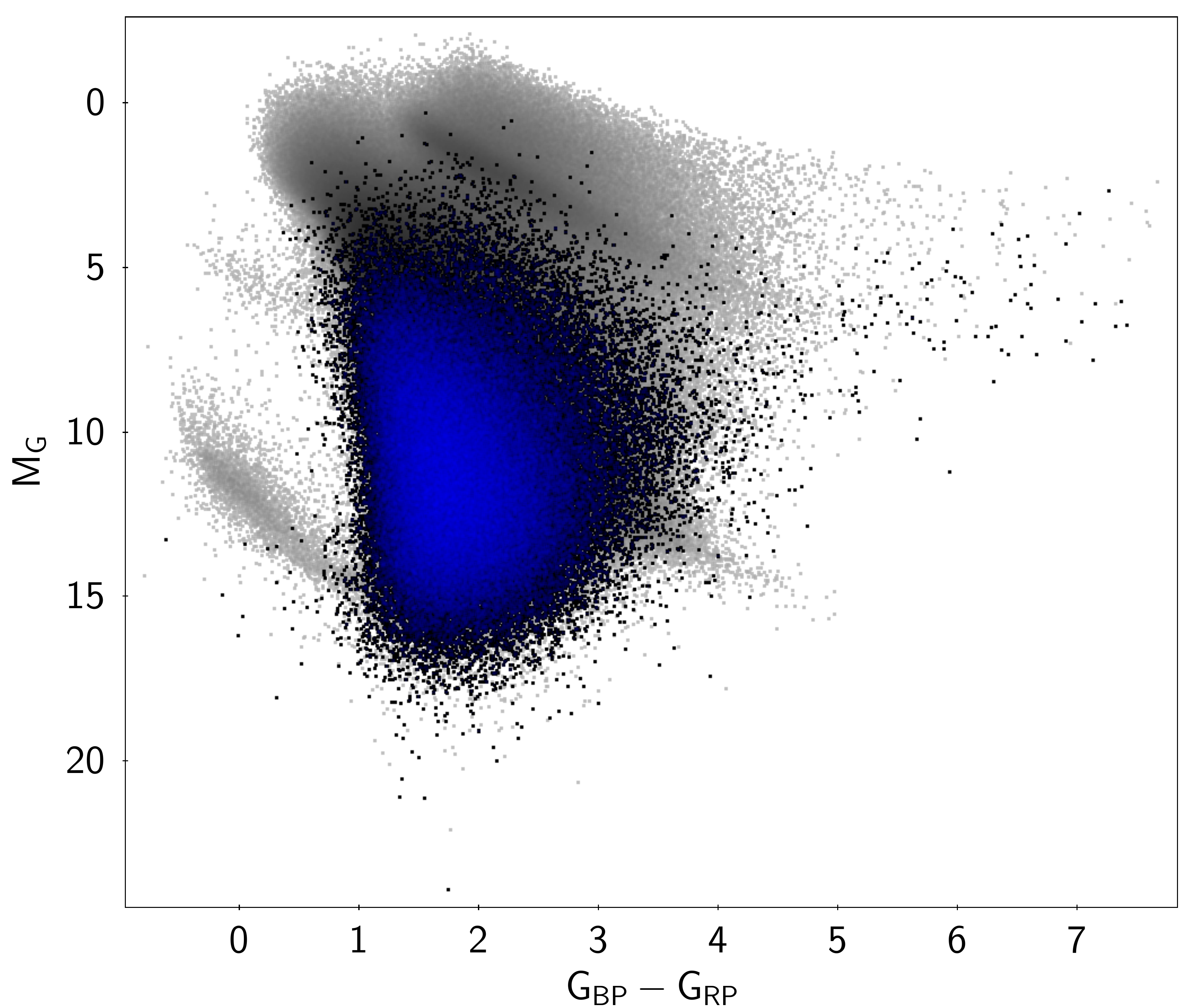}
    \caption{The actual \gaia/IPHAS catalogue is shown with grey points, while our negative parallax ``mirror sample'' -- which highlights the regions where we can expect false positive to occur -- are shown in blue/black. Left panel: $G$-band magnitude versus reduced $\chi^2$ ($\chi_{\nu}^2$, based on the \gaia\ astrometric fit). Right panel: \gaia-based CMD for the same targets. Although not realistic, we have adopted absolute values of the parallax measurements to infer the absolute magnitudes of the ``mirror'' negative parallax sample to compare their position in the CMD to those ``problematic sources'' (\citealt{gaiaAstro}).}
    \label{fig:IPHAS_PosNeg}
\end{figure*}

The left panel of Figure \ref{fig:IPHAS_PosNeg} shows the ($G$,$\chi_{\nu}^2$) plane for targets in our catalogue, together with our negative parallax mirror sample overlayed. The right panel shows the resulting \gaia\ CMD for the our full catalogue with our mirror sample overlayed, where we have used the absolute parallax value to infer absolute magnitude. The assumption made in adopting absolute parallax values for negative parallax targets is that the \gaia\ DR2 processing occasionally produces spurious astrometry that may equally well result in a positive or negative parallax. This assumption is tested using the absolute value. The similarity in the location of the ``blob'' in CMD space to where most targets are expected not to be reliable is remarkable. Also striking is the similarity between the region in CMD space occupied by our miss-matched sample (Figure \ref{fig:GaiaIPHAS_CMD_BAD}, right-panel) and our mirror sample, confirming our assumption. 

Using both our catalogue and the mirror sample we define a new quality parameter for each object in our catalogue which identifies the probability for a particular object to be a false-positive entry based on its position in the $(G,\chi_{\nu}^2)$ plane. We do this by first binning all targets (including our mirror sample) in $G$-band bins of 0.1 magnitudes, with the additional requirement that each bin contains at least 10,000 objects, similarly to what has been done in Section \ref{sec:completeness}. For each $G$-band bin, we then sort all entries by increasing $\chi_{\nu}^2$, and subsequently bin these into blocks of 1000 objects. For each target in our catalogue which lies in a particular block we then define the false-positive fraction ($f_{FP}$) as the number of objects in the mirror sample in this block, divided by the total number of objects in the block (i.e. $f_{FP} = \frac{N_{neg}}{N_{pos} + N_{neg}}$, where $N_{neg}$ and $N_{pos}$ are the number of negative and positive parallax objects within a specific block). It is important to realise that our definition of $f_{FP}$ ensures that the obtained values will strictly be within the $0\leq f_{FP}<1$ range. In the low $f_{FP}$ regime, applicable to the \gaia\ dataset, our definition can be interpreted as $f_{FP} = \frac{N_{neg}}{N_{pos} + N_{neg}} \approx \frac{N_{neg}}{N_{pos}}$.

\begin{figure}
	\includegraphics[width=1\columnwidth]{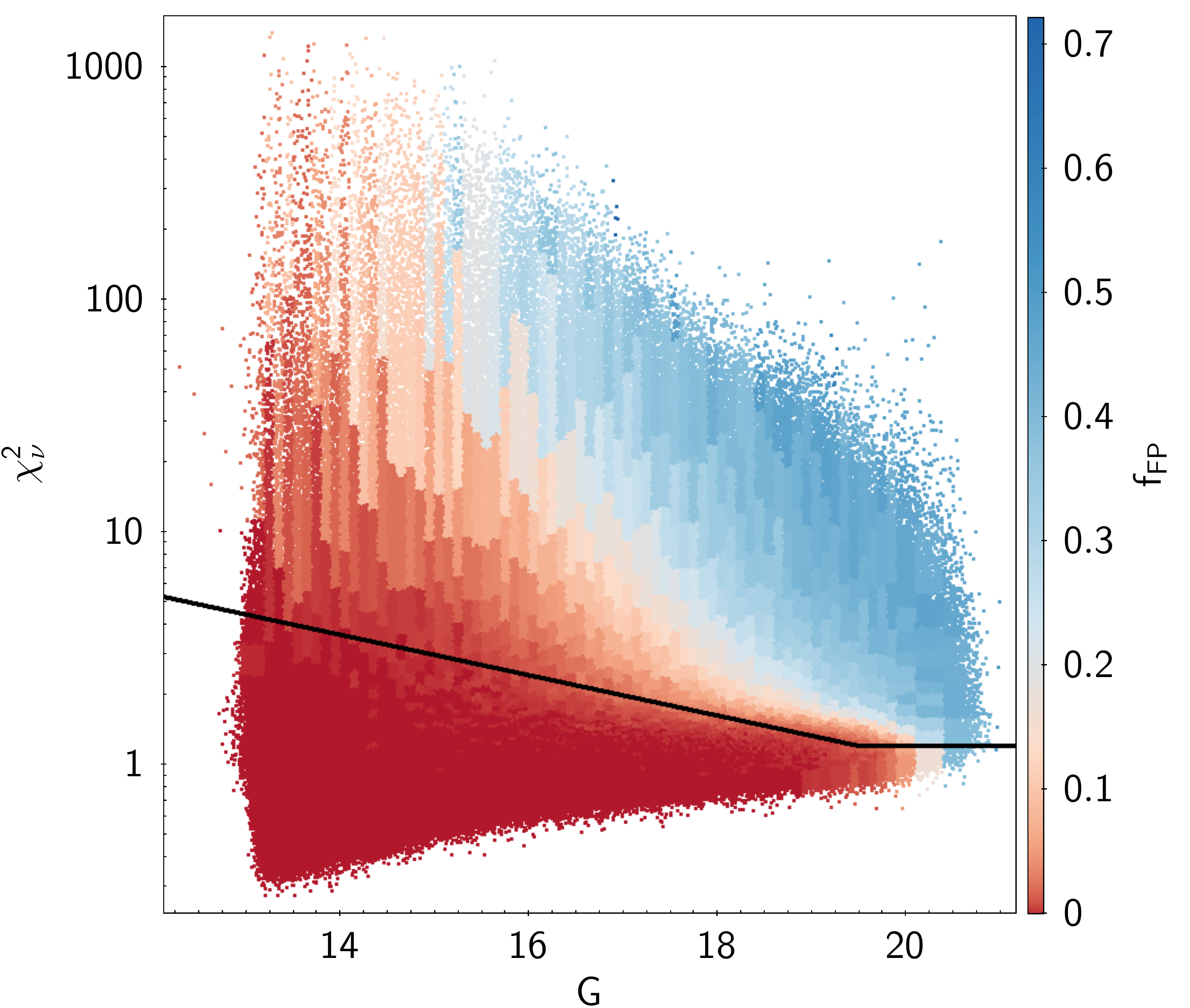}
    \caption{\gaia\ $G$-band magnitude versus reduced $\chi^2$ ($\chi_{\nu}^2$, based on the \gaia\ astrometric fit) for all targets retained in our \gaia/IPHAS catalogue. Targets are colour coded by their false-positive rate fraction $f_{FP}$, based on the number density of targets with negative parallaxes occupying a specific region in the ($G$,$\chi_{\nu}^2$) parameter space (see Section \ref{sec:completeness} for details). The thick black line shows the recommended cut discussed in the Appendix of \citet{gaiaAstro} (Equation C.1)}
    \label{fig:Gchi2_fFP}
\end{figure}

Figure \ref{fig:Gchi2_fFP} shows the $(G,\chi_{\nu}^2)$ plane colour-coded with $f_{FP}$. Similarly to Figure \ref{fig:Gchi2_fc}, the apparent ``edges'' in $f_{FP}$ are caused by the set limits on the bin/block sizes and/or number of objects per bin/block. This method can potentially be applied to other, larger, catalogues where this effect can be minimised with smaller bins/blocks. Furthermore, given the requirement of 1,000 objects per block, the precision with which we can determine $f_{FP}$ is limited to 0.001.

\subsection{Cleaning the \gaia/IPHAS and \gaia/KIS CMDs} \label{sec:clean}
Having included for each target in our catalogues two additional quality parameters, $f_c$ and $f_{FP}$, we can now use these to clean the catalogues to produce more reliable sets of targets. 

\cite{arenou18} shows that spurious astrometric solutions are more common in specific areas of the sky, and may depend on both the \gaia\ scan directions and epoch of observation. Additionally, spurious astrometric solutions are more frequent in dense areas of the sky, particularly the Galactic plane. Our approach of computing $f_c$ and $f_{FP}$ is independent of sky position, and it may be that these values differ depending on sky position. However, as IPHAS is concentrated on the Galactic plane where most of the \gaia\ spurious measurements are found, these differences should be relatively small. We also expect the $f_{FP}$ fraction to become more relaxed for other sky areas out of the Galactic plane.

Ideally, if the only problem with our catalogues was the sort of statistical error that is responsible for the existence of the mirror negative parallax sample, there would be no need to have the additional $f_c$ parameter. However, in practice, the sole use of $f_{FP}$ does not remove all sources with bad astrometry. It is therefore useful to use both $f_c$ and $f_{FP}$ to clean our catalogues. We note that when both quality cuts are employed, the ``completeness'' of a subset can no longer be guaranteed to be greater than $f_c$, since it might occur that within specific $G$-band magnitude bins the $f_{FP}$ threshold will remove additional objects.

\begin{figure*}
	\includegraphics[width=0.66\columnwidth]{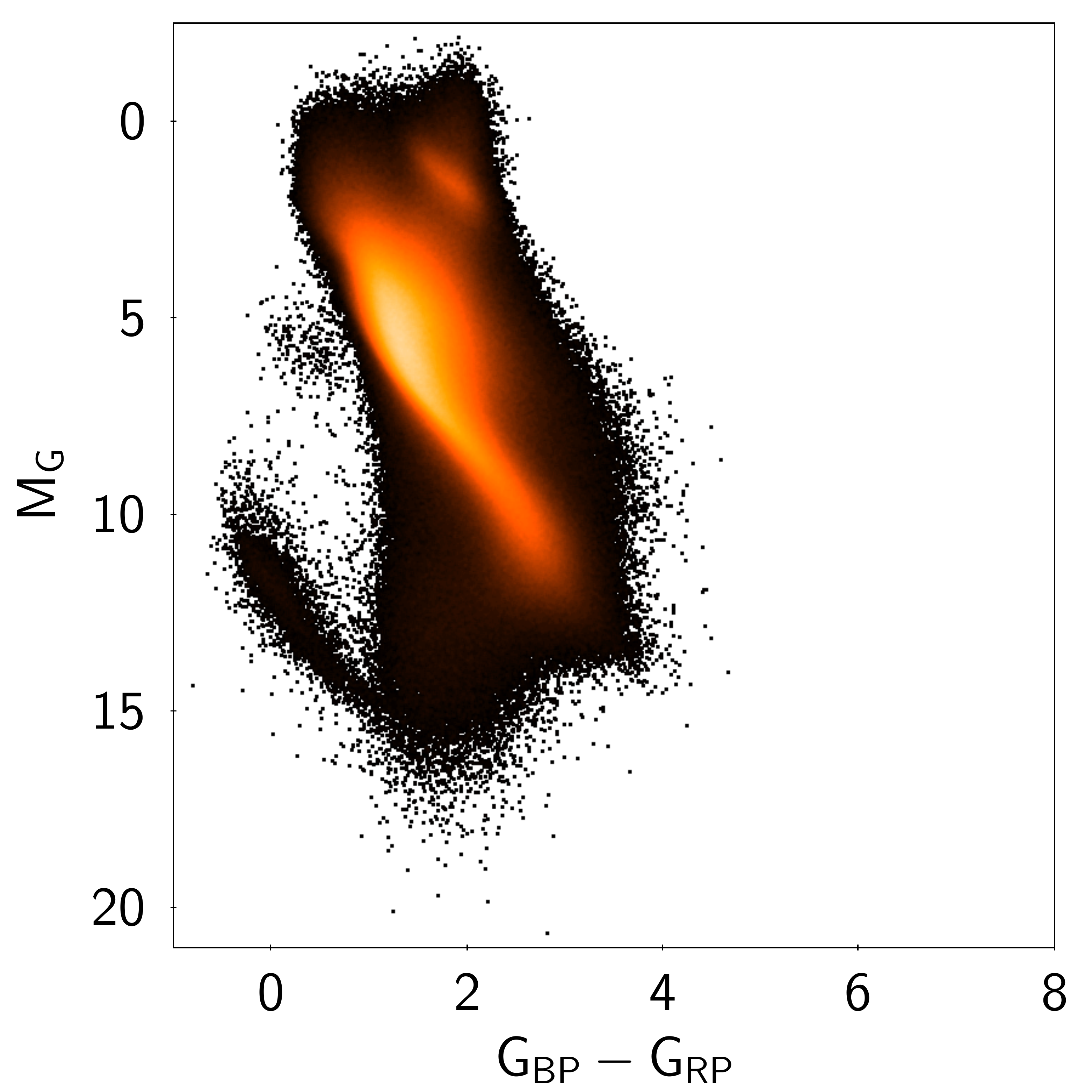}
    \includegraphics[width=0.66\columnwidth]{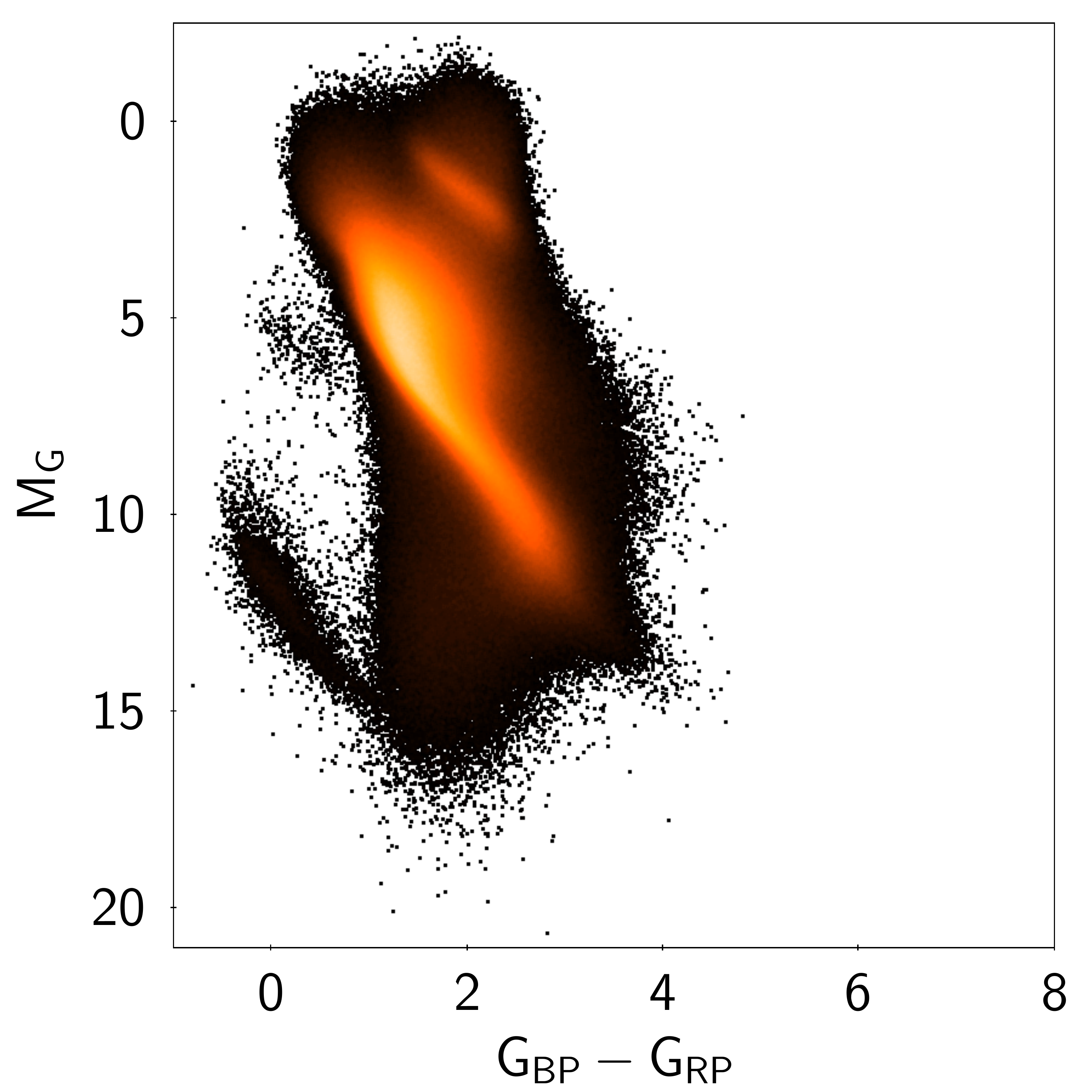}
    \includegraphics[width=0.66\columnwidth]{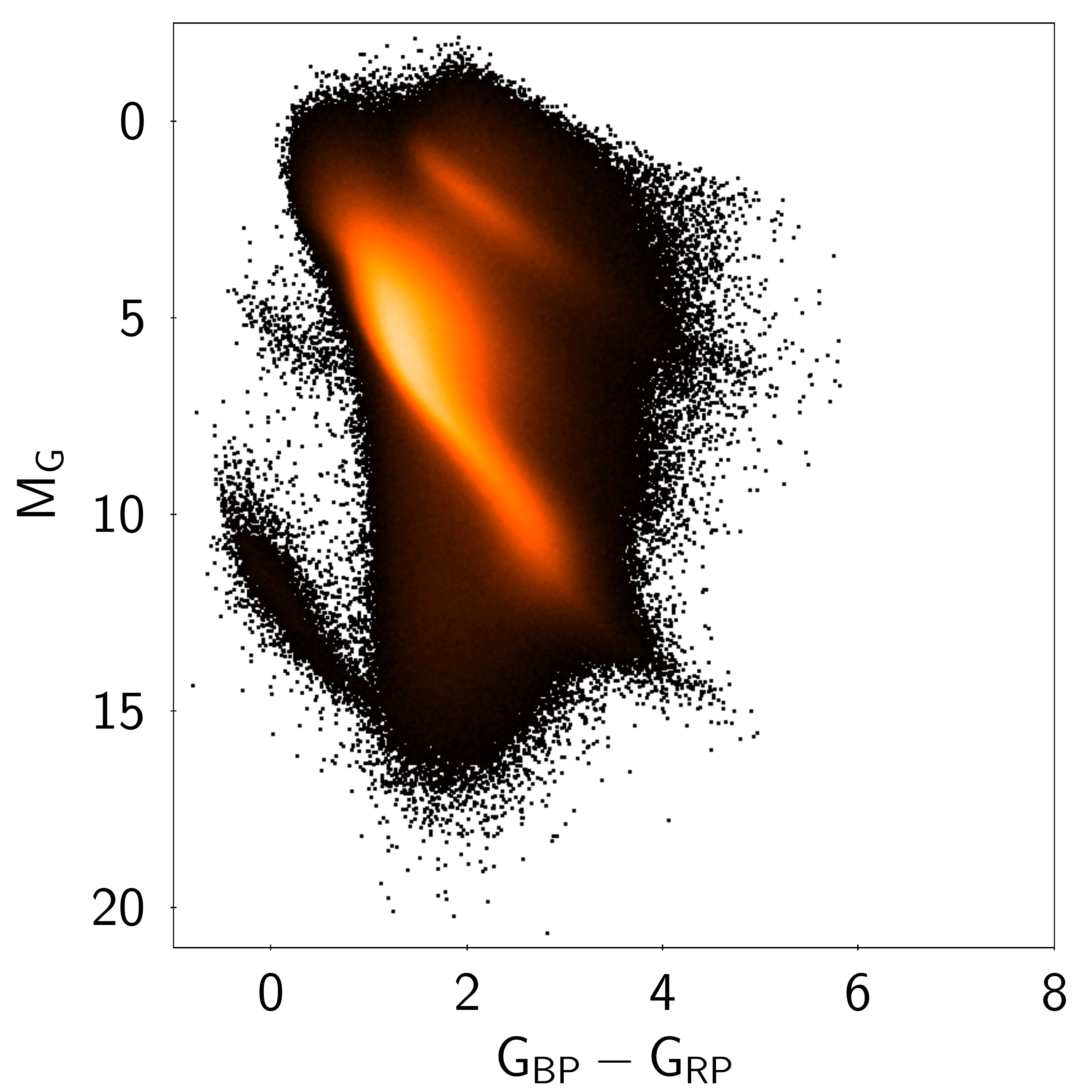}
    \newline
    \includegraphics[width=0.66\columnwidth]{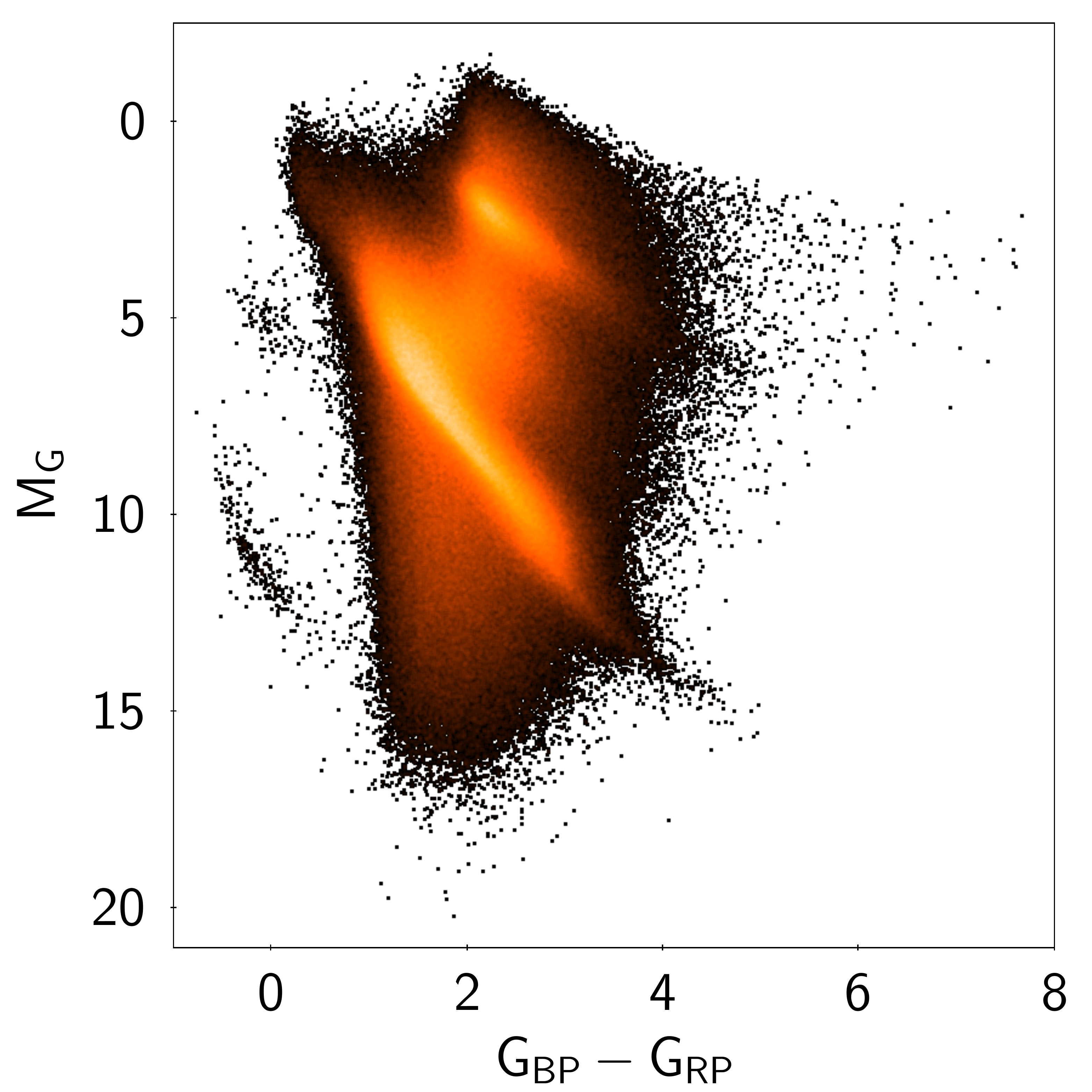}
    \includegraphics[width=0.66\columnwidth]{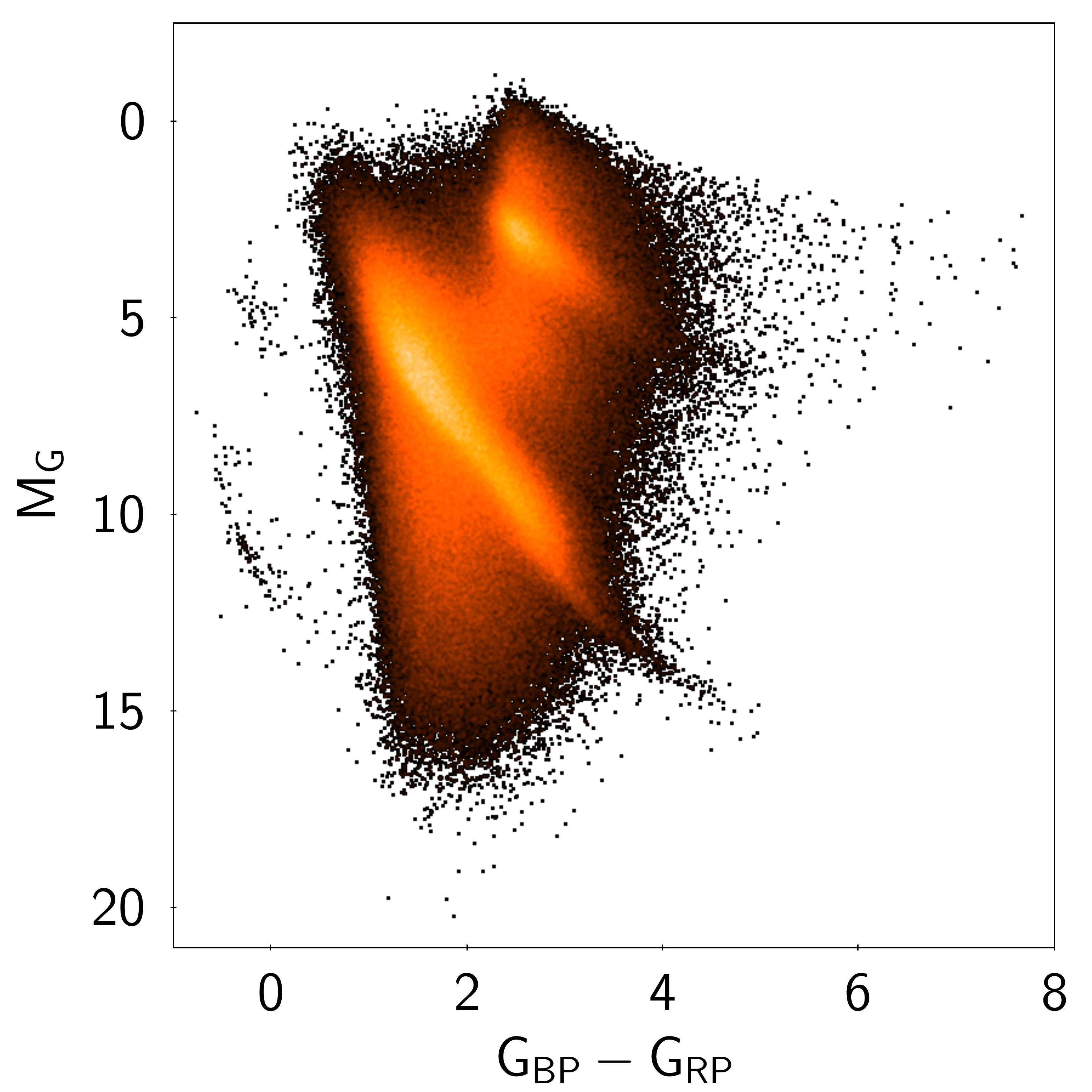}
    \includegraphics[width=0.66\columnwidth]{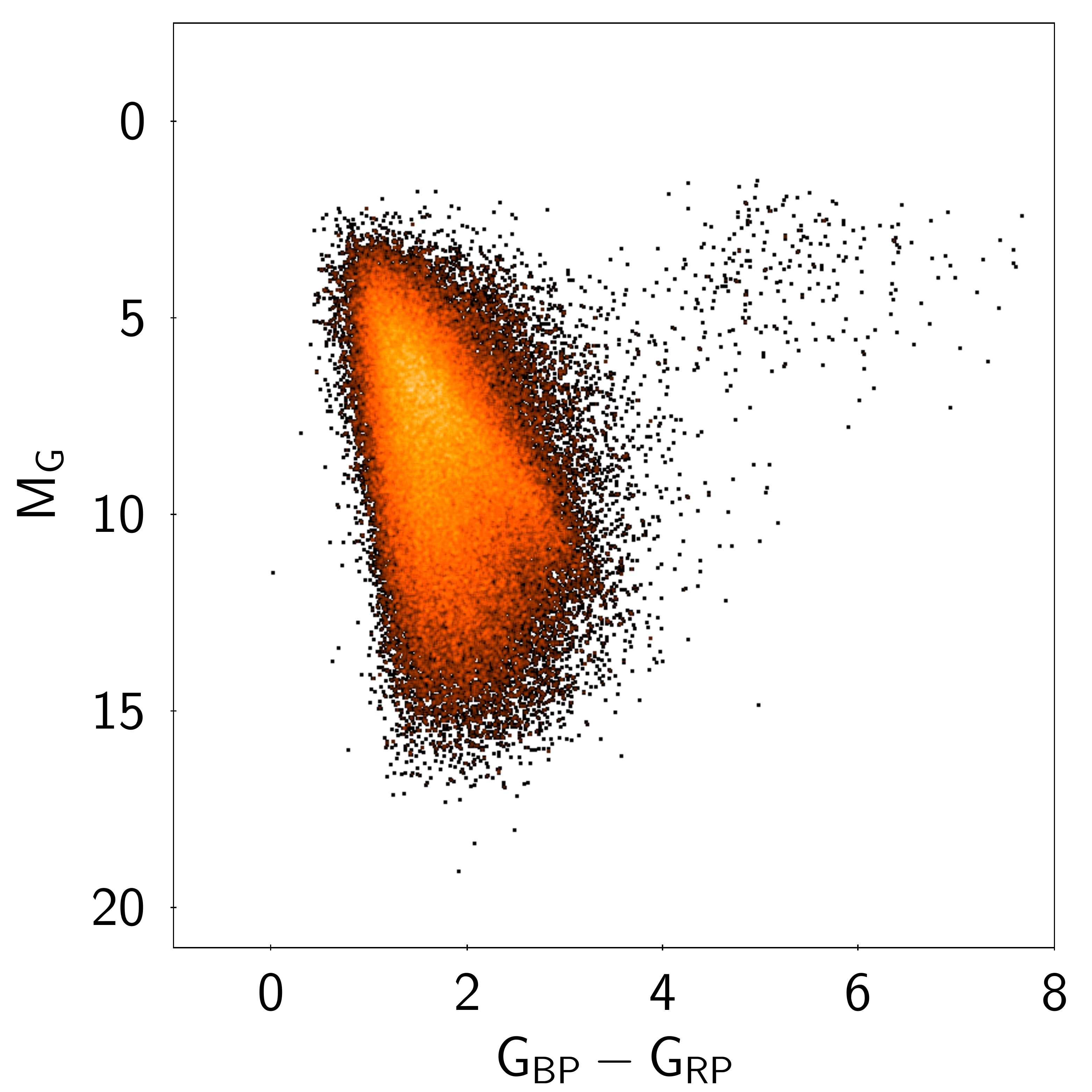}
    \caption{\gaia\ CMDs of our \gaia/IPHAS catalogue with varying threshold cuts on our defined ``completeness'' parameter ($f_c$). From left to right the cuts employed are $f_c<0.8$, $f_c<0.9$ and $f_c<0.99$. The top panels show the retained sources adopting a specific cut, whilst the bottom panels show the removed sources.}
    \label{fig:Gaia_CMD_fc}
\end{figure*}

\begin{figure*}
	\includegraphics[width=0.66\columnwidth]{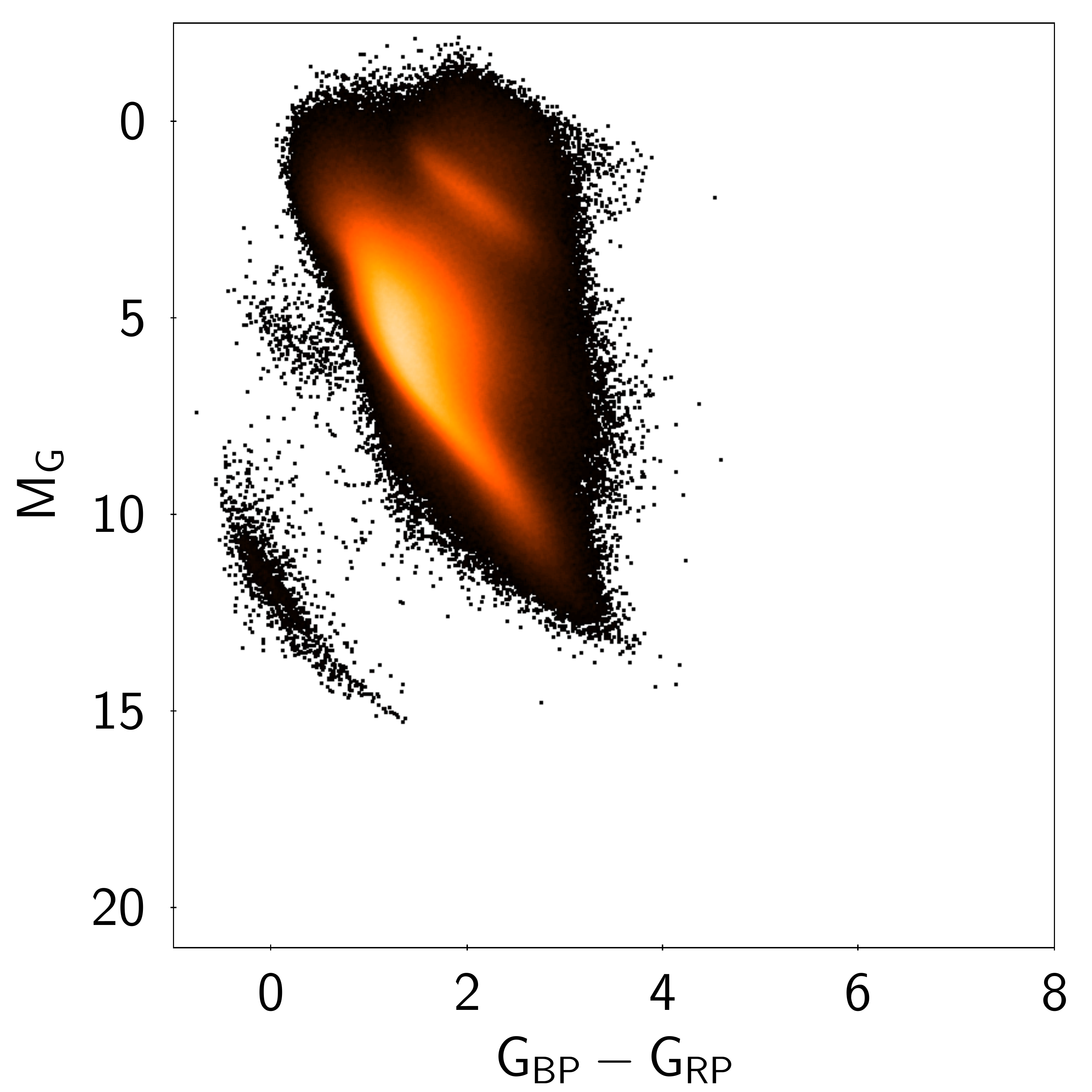}
    \includegraphics[width=0.66\columnwidth]{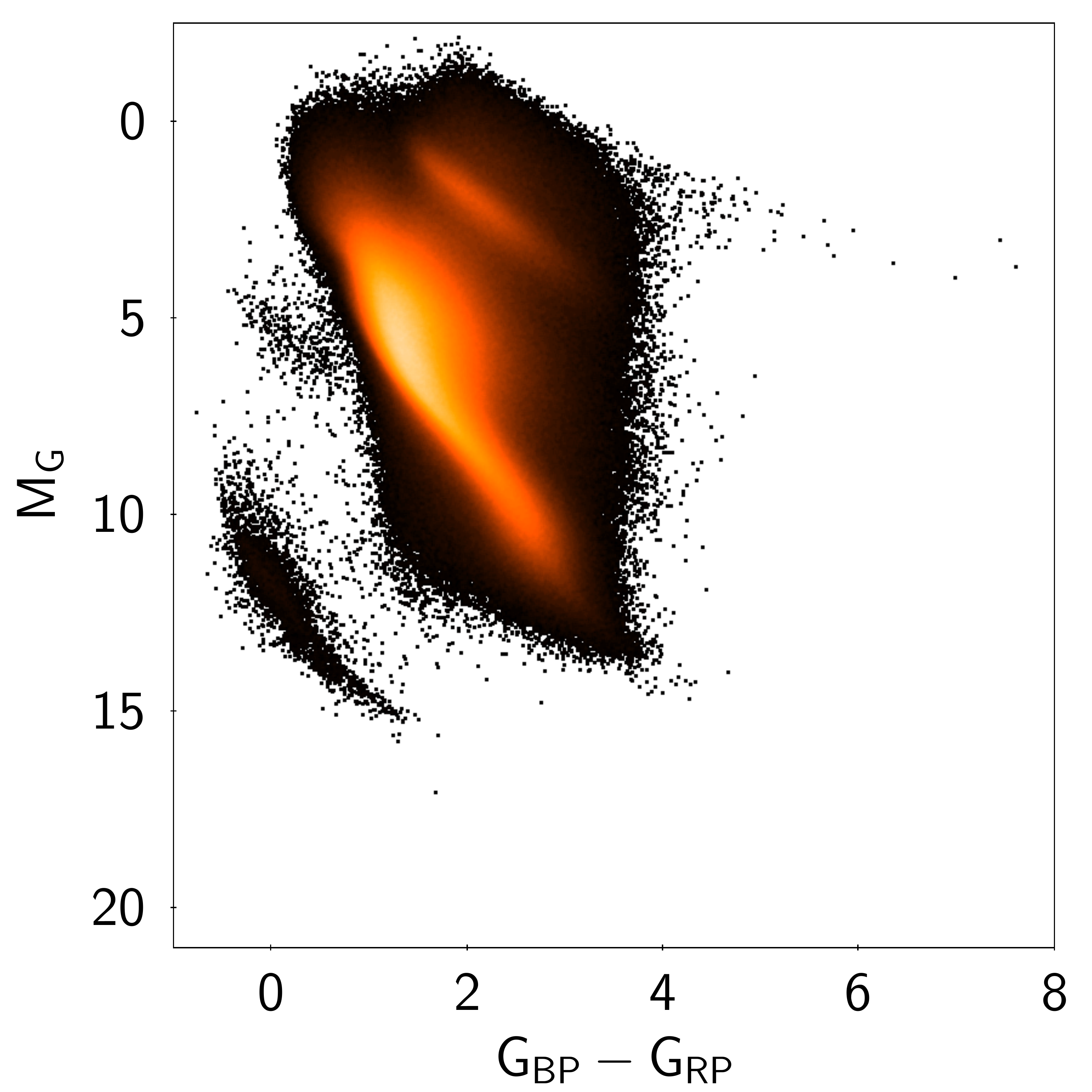}
    \includegraphics[width=0.66\columnwidth]{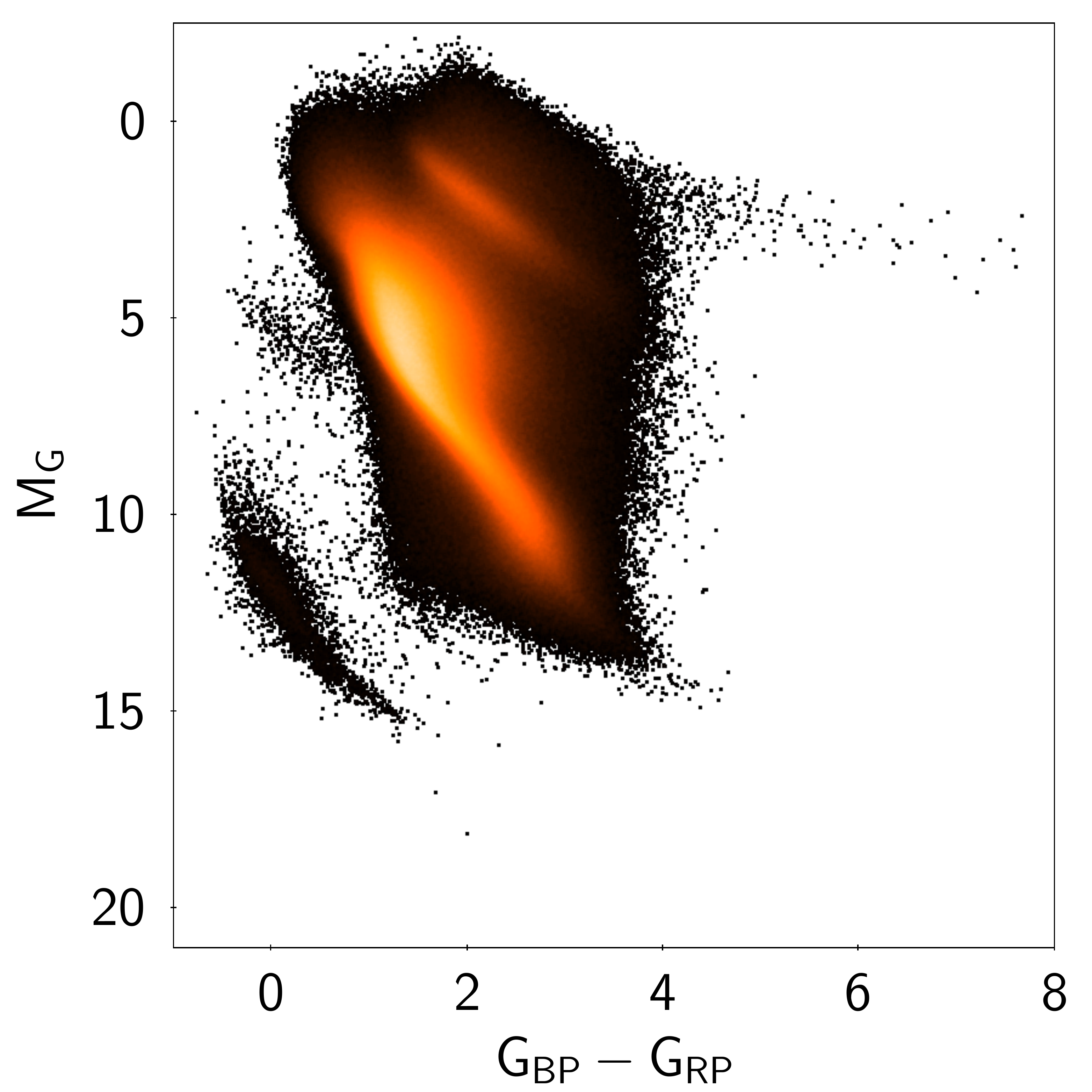}
    \newline
    \includegraphics[width=0.66\columnwidth]{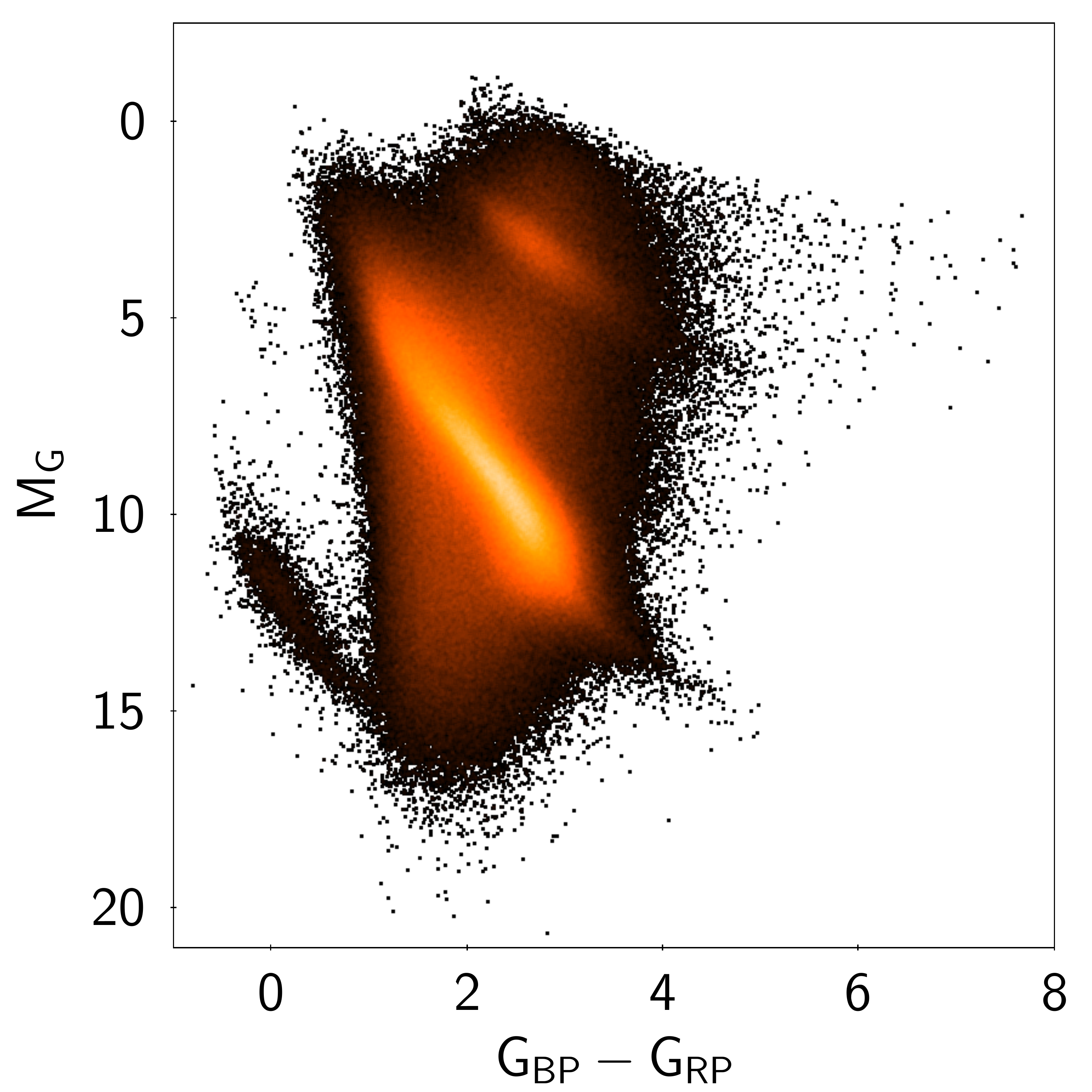}
    \includegraphics[width=0.66\columnwidth]{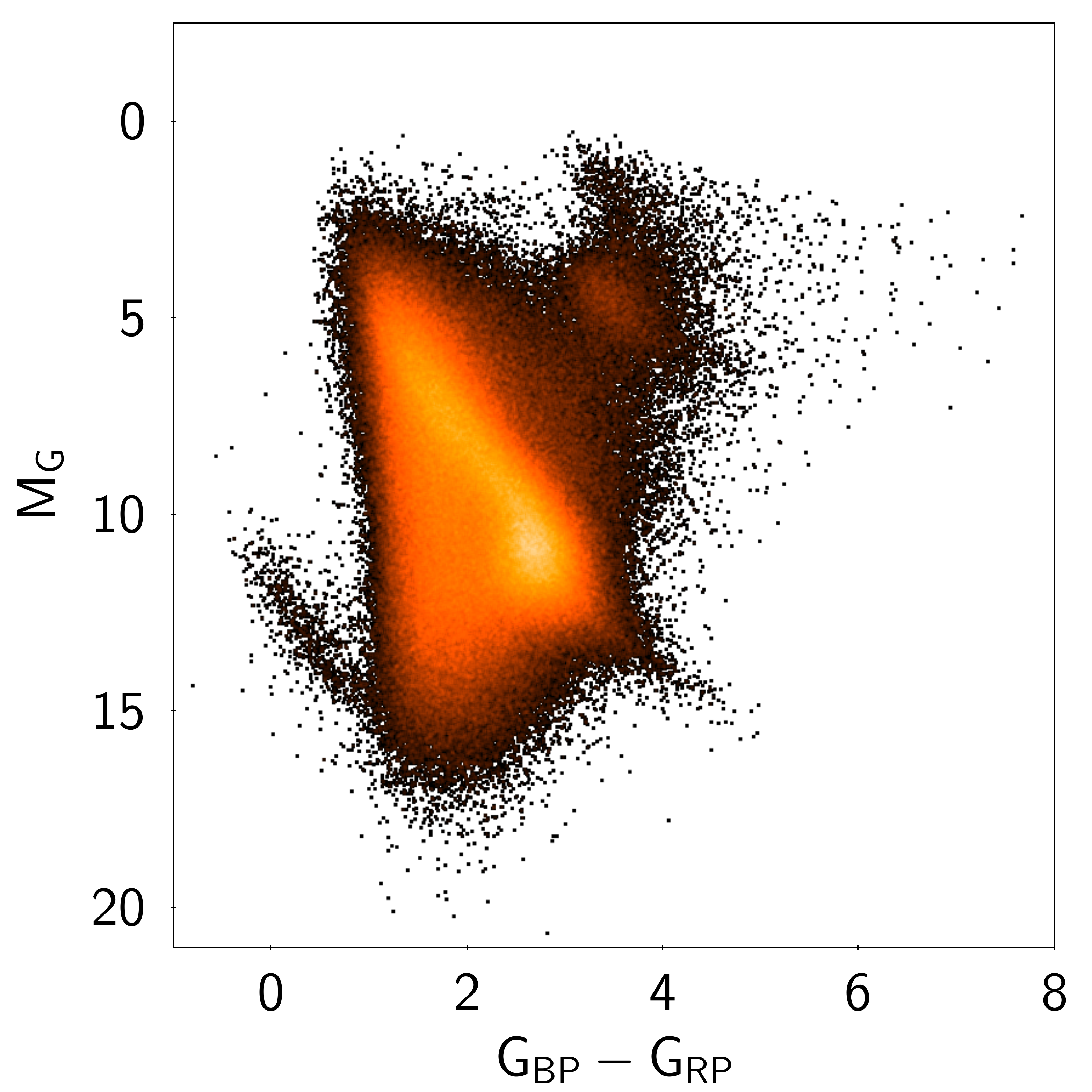}
    \includegraphics[width=0.66\columnwidth]{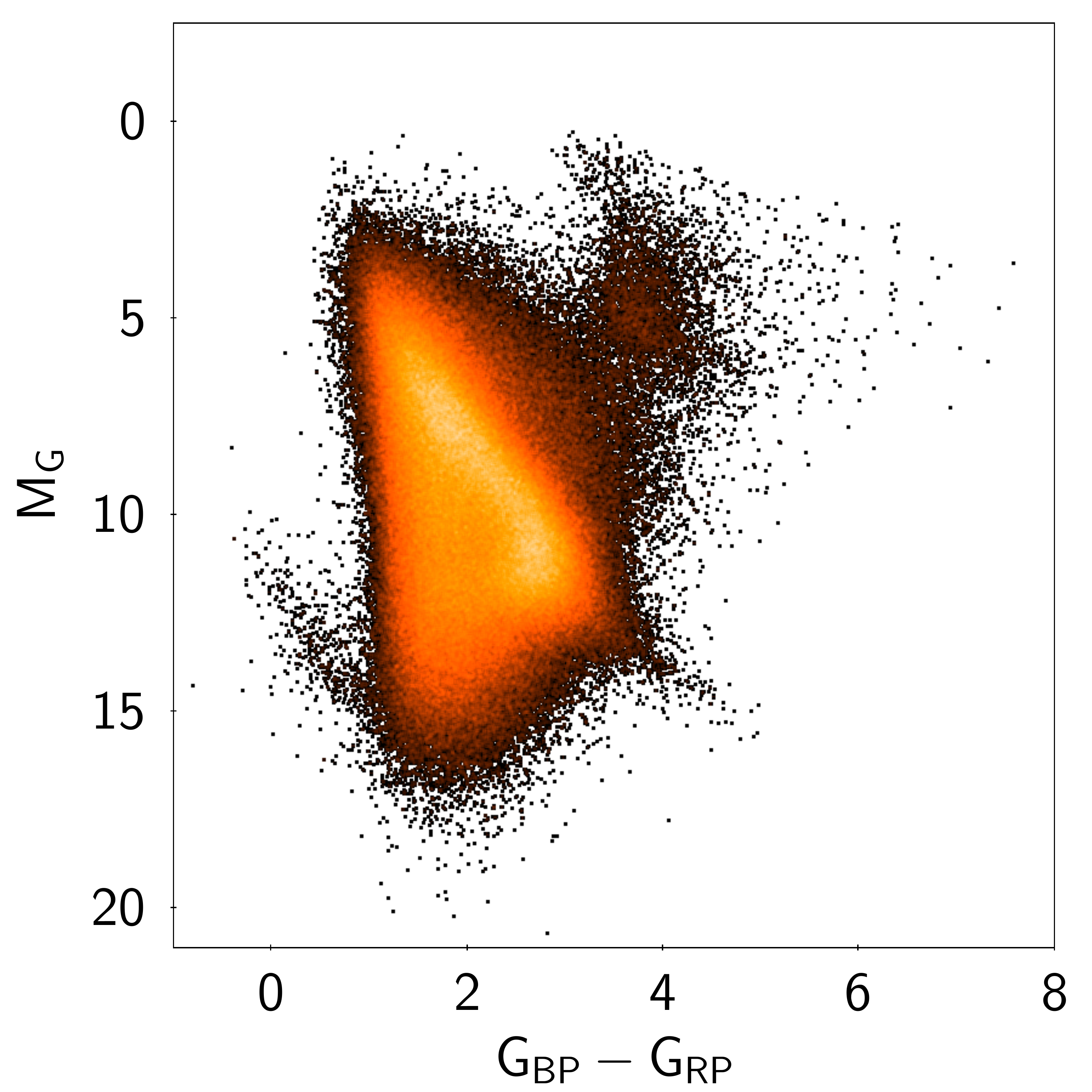}
    \caption{\gaia\ CMDs of our \gaia/IPHAS catalogue with varying threshold cuts on our defined false-positive parameter ($f_{FP}$). From left to right the cuts employed are $f_{FP}=0$, $f_{FP}\leq 0.01$ and $f_{FP}\leq 0.02$. The top panels show the retained sources adopting a specific cut, whilst the bottom panels show the removed sources.}
    \label{fig:Gaia_CMD_fFP}
\end{figure*}

To illustrate the effect of selecting targets based on our quality parameters, we show in Figure \ref{fig:Gaia_CMD_fc} and Figure \ref{fig:Gaia_CMD_fFP} the \gaia\ CMDs for varying $f_c$ and $f_{FP}$ thresholds, respectively. The top panels in both Figures show the retained targets, whilst the bottom panels show the removed ones.

\begin{figure*}
	\includegraphics[width=0.66\columnwidth]{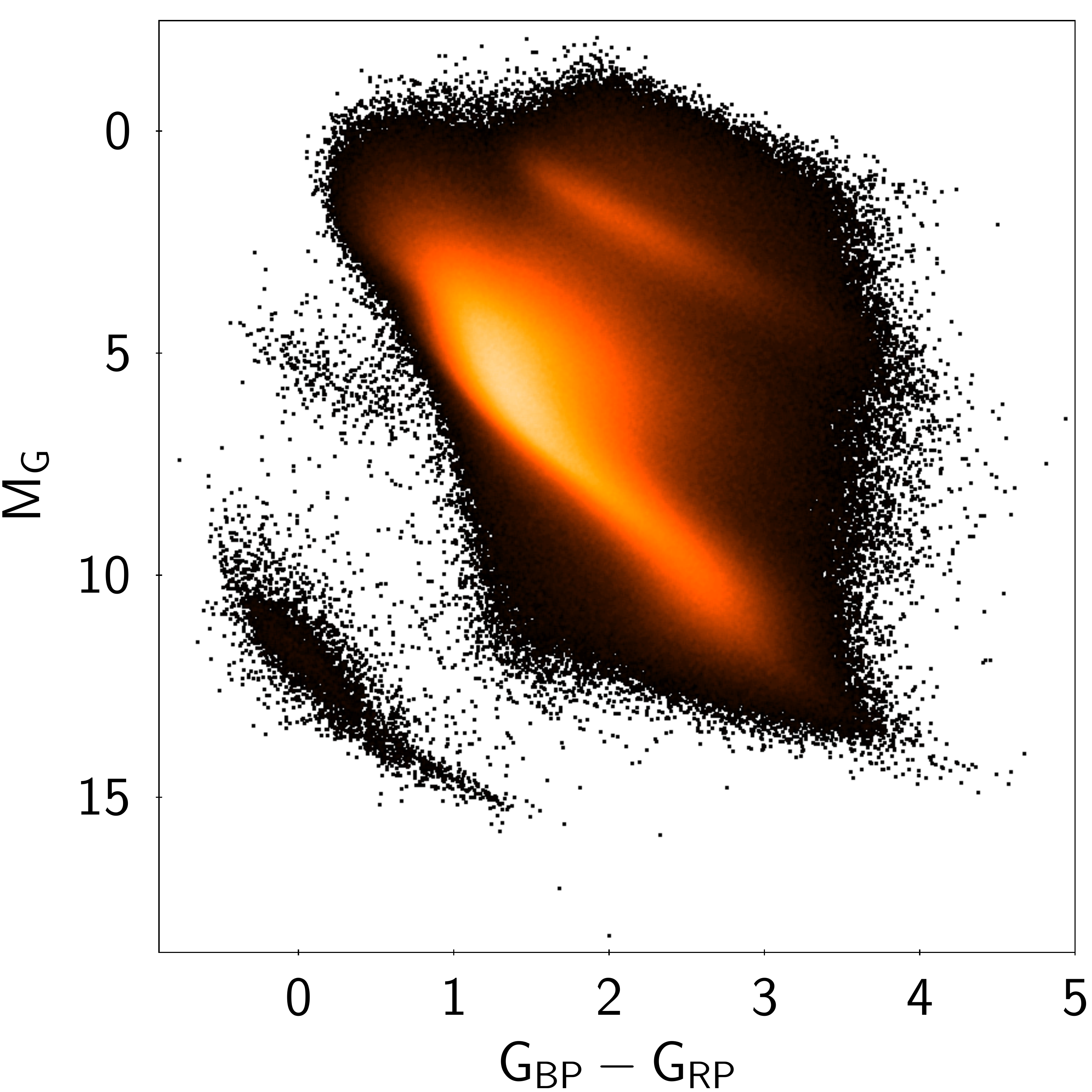}
    \includegraphics[width=0.66\columnwidth]{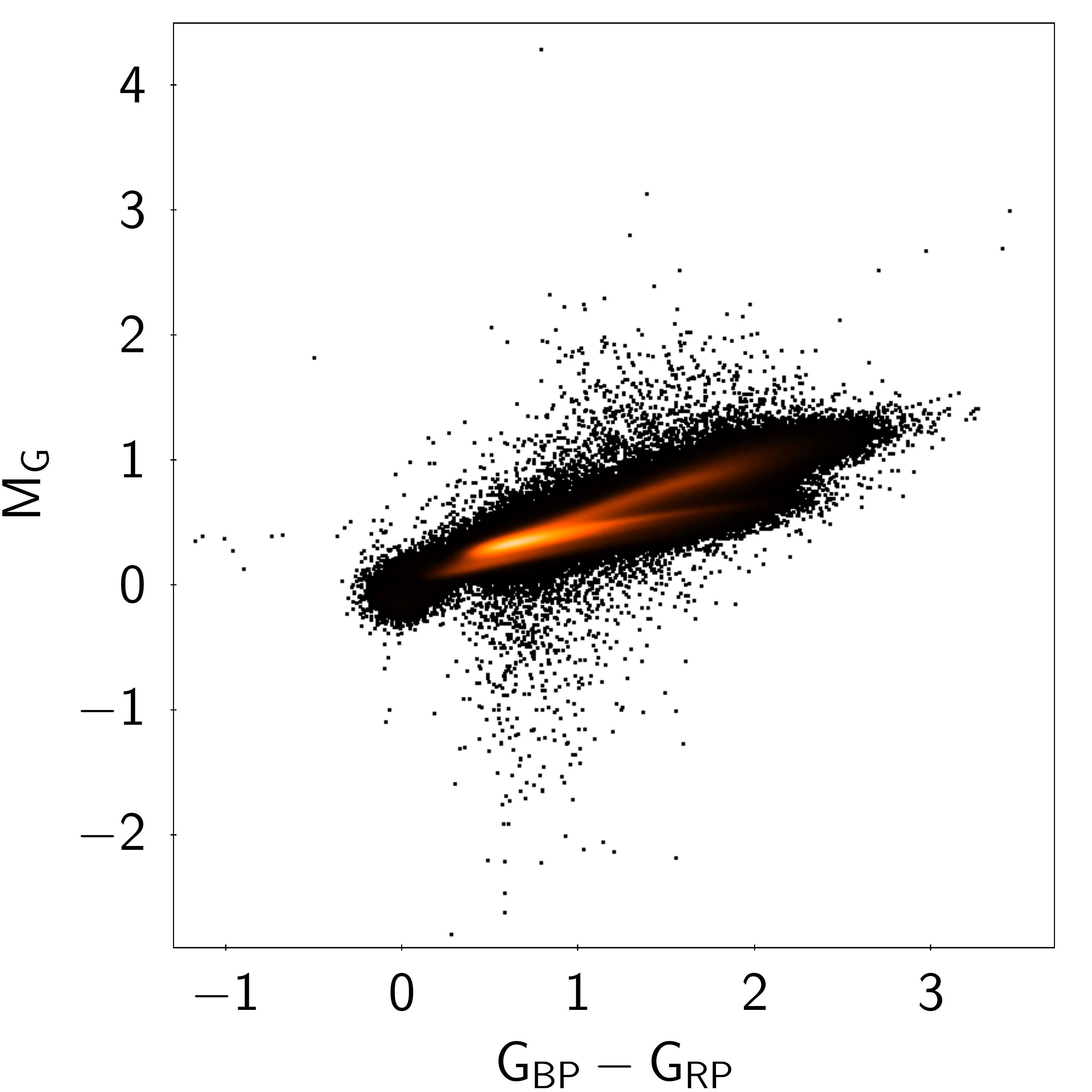}
    \includegraphics[width=0.66\columnwidth]{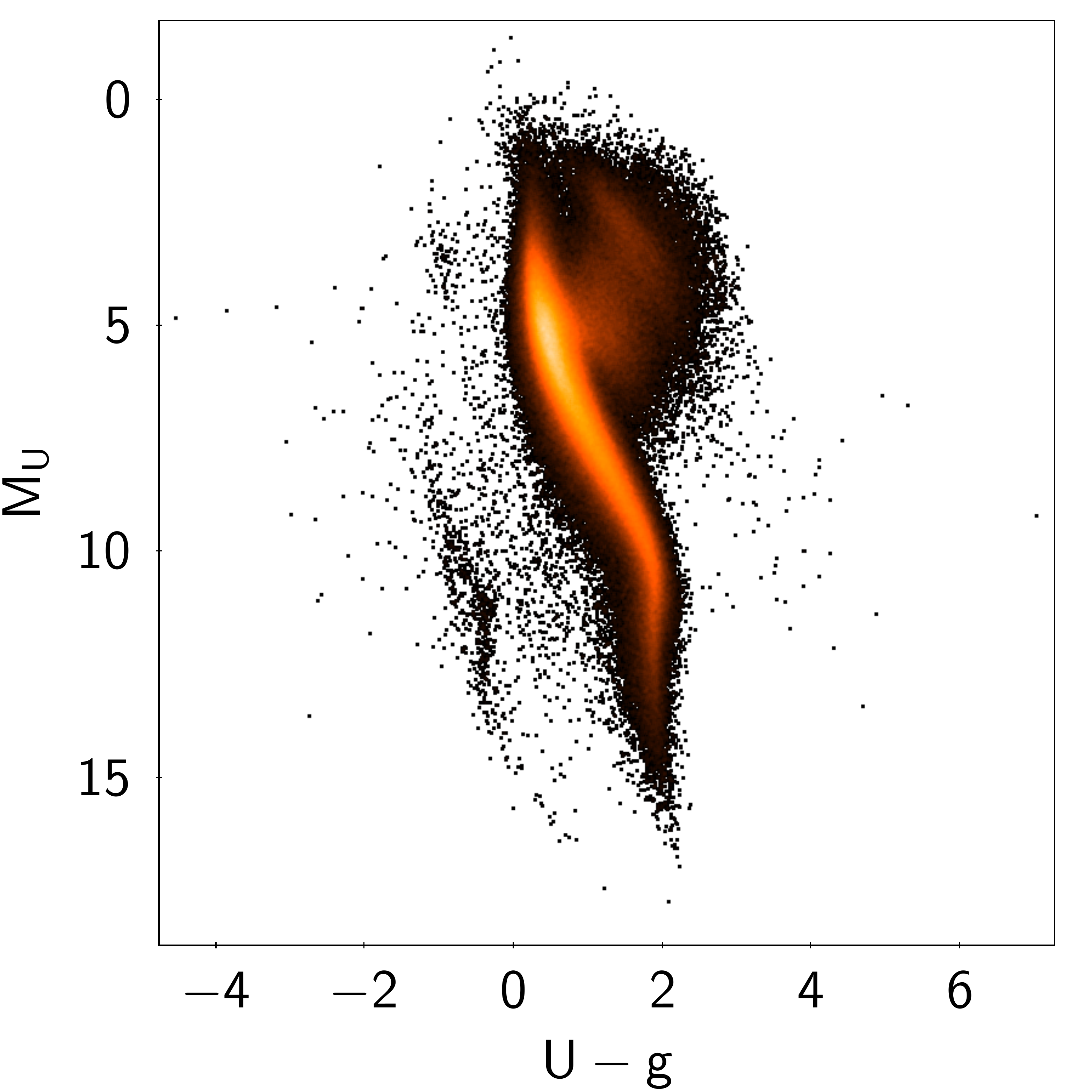}
    \caption{CMDs and colour-colour diagrams from targets retained after adopting our recommended quality cuts on $f_c$ and $f_{FP}$. Left panel: \gaia-based CMD from our \gaia/IPHAS catalogue. Middle panel: IPHAS colours from our \gaia/IPHAS catalogue. Right panel: KIS-based CMD from our \gaia/KIS catalogue.}
    \label{fig:bestCMD}
\end{figure*}

Based on visually inspecting various CMDs with all of the \gaia, IPHAS, and KIS photometry, our recommended general-purpose quality cuts are $f_c < 0.98$ and $f_{FP} \leq 0.02$. Using these thresholds retains 94\% and 98\% from our \gaia/IPHAS catalogue and \gaia/KIS catalogues respectively. These cuts provide a relatively clean subset of objects, with minimal false-positives and a large retention fraction. Figure \ref{fig:bestCMD} shows some of the CMDs and colour-colour plots produced using our recommended quality cuts for both the IPHAS and KIS merged-catalogues with \gaia. 

\begin{figure*}
	\includegraphics[width=1\columnwidth]{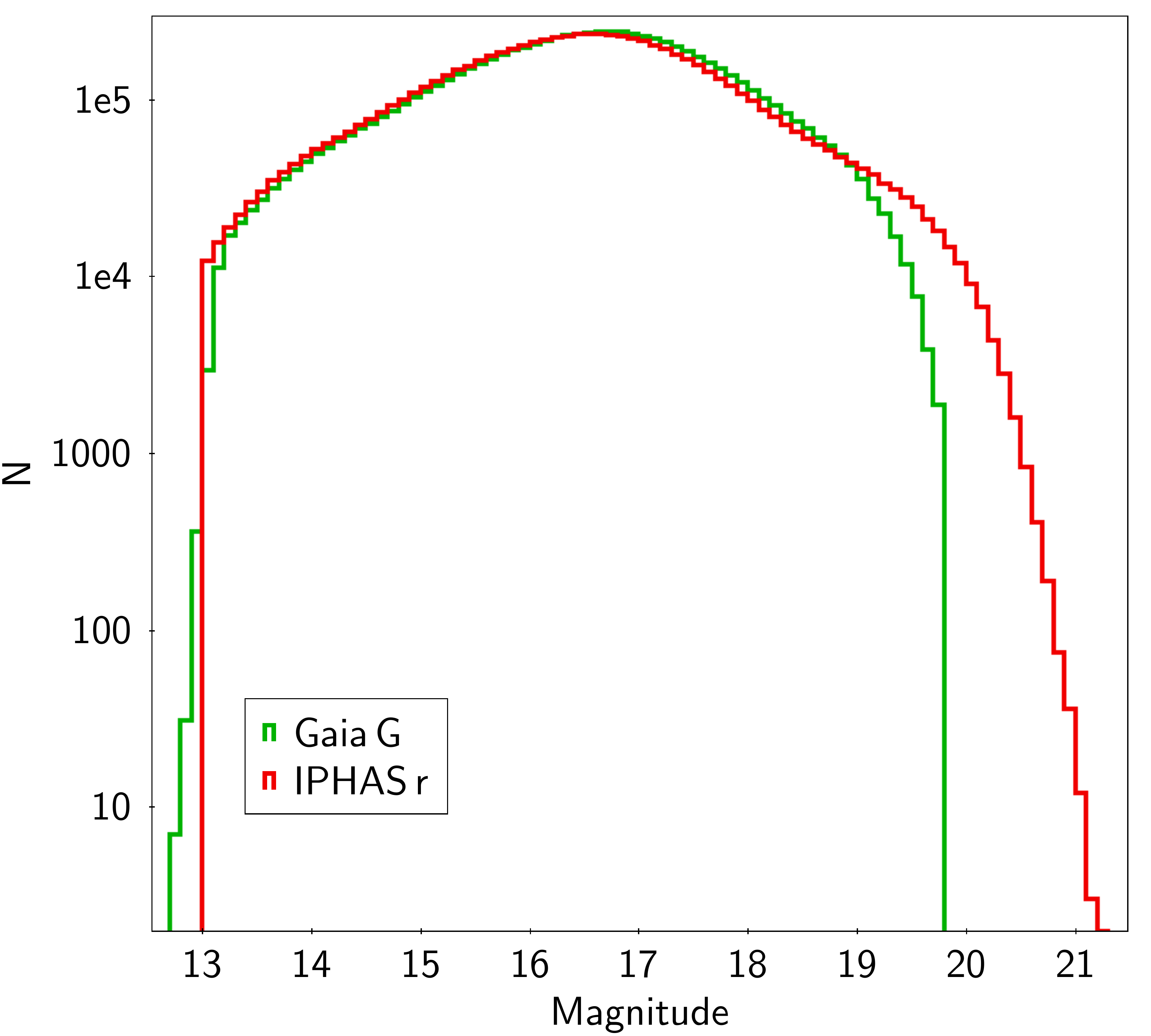}
    \includegraphics[width=1\columnwidth]{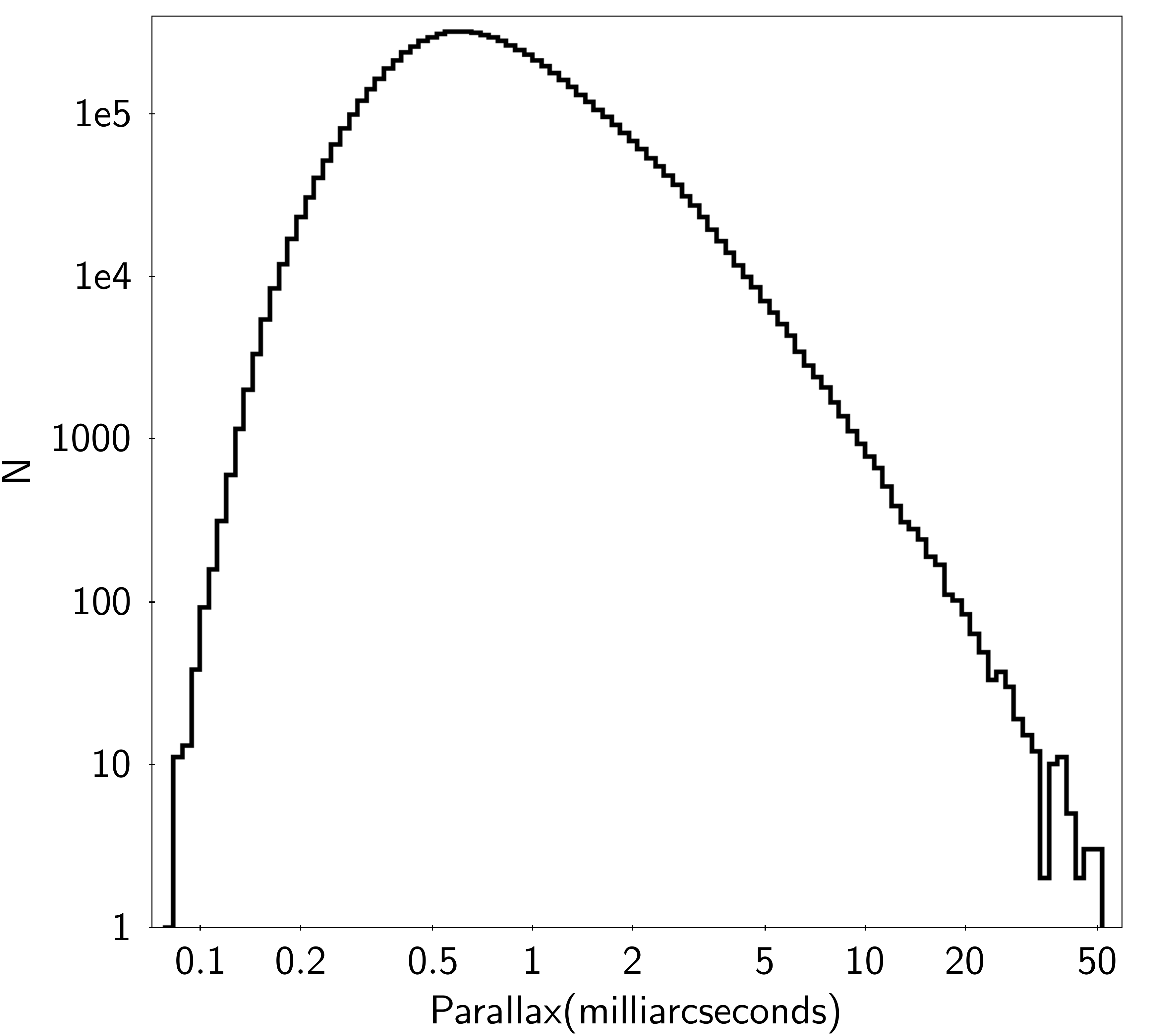}
    \caption{$r$ and $G$ band magnitude distributions (left panel) and parallax distribution (right panel) from the \gaia/IPHAS catalogue for all objects satisfying our quality control cuts discussed in Section \ref{sec:clean}.}
    \label{fig:dists}
\end{figure*}

Figure \ref{fig:dists} shows the $r$, $G$ and parallax distributions for the \gaia/IPHAS catalogue adopting our recommended quality cuts. From these distributions we can comment that the cross-matched \gaia/IPHAS catalogue becomes incomplete for source fainter than $\simeq$ 16-17 mag, which expressed in terms of distance is growing incompleteness between 1 and 2 kpc. In directions of high extinction such as towards the Aquila and Cygnus Rifts, the distance turnover will be closer than elsewhere. This property is primarily a consequence of the \texttt{parallax\_over\_error}>5 selection criterion (see Section \ref{sec:selection}).

\section{Enabling additional science exploitation with Gaia and IPHAS/KIS} \label{sec:science}
In this section we will highlight some of the possible science exploitations possible with the value-added \gaia/IPHAS and \gaia/KIS catalogues. As in the previous section we will take examples from the \gaia/IPHAS catalogue, but the same exercises can also be performed with the \gaia/KIS catalogue.

\subsection{Identifying high proper motion objects}
Because of the sub-arcsecond cross-matching precision between \gaia\ and IPHAS/KIS, it is now possible to gather additional photometric information for some of the highest proper motion objects.

\begin{figure}
	\includegraphics[width=1\columnwidth]{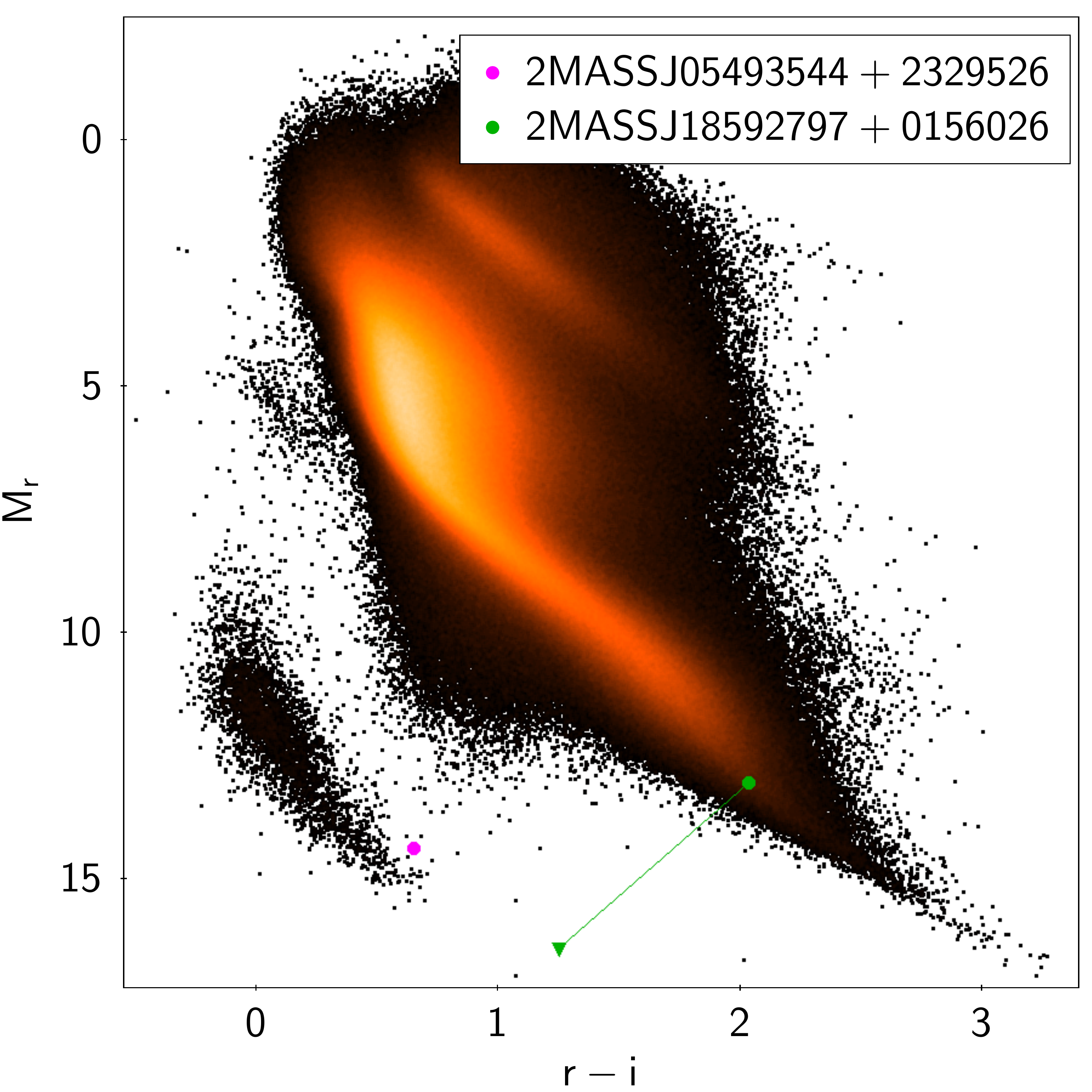}
    \caption{IPHAS-based CMD using our recommended quality cuts showing the positions of two example sources which would have either been missed (magenta point) entirely or miss-matched (green point). The green line connecting the green triangle to the green point demonstrates the change in CMD position for the miss-matched object in question. The IPHAS images for these targets are shown in Figure \ref{fig:missmatchGaia}.}
    \label{fig:GaiaCleanPM}
\end{figure}

\begin{figure*}
	\includegraphics[width=2\columnwidth]{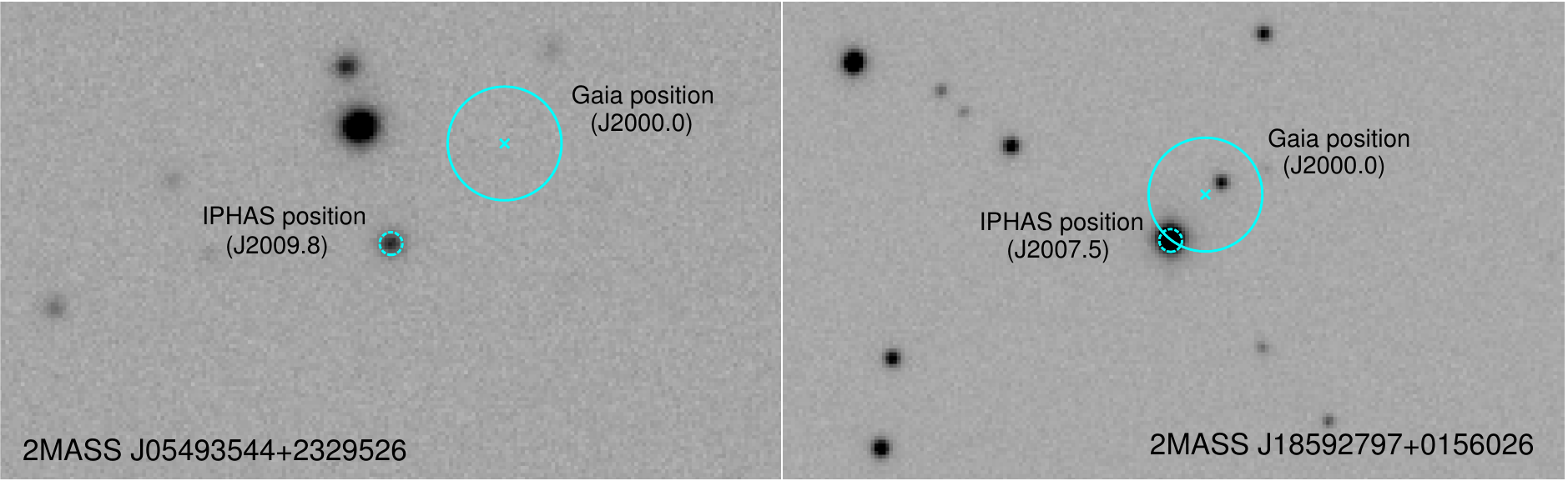}
    \caption{IPHAS $r$-band image of the recovered target 2MASSJ05493544$+$2329526 (left) and 2MASS J18592797$+$0156026 (right). The dashed 1'' circles are centred on the recorded IPHAS positions. The solid 5'' circles are centred on the rewinded J2000.0 \gaia\ positions of the same target. The positions in the IPHAS CMD for these targets are shown in Figure \ref{fig:GaiaCleanPM}.}
    \label{fig:missmatchGaia}
\end{figure*}

Figure \ref{fig:GaiaCleanPM} shows our cleaned \gaia/IPHAS sources in the absolute IPHAS CMD diagram using the distances inferred from the \gaia\ parallaxes. Two objects are marked for illustration purposes, both also appearing in Figure \ref{fig:GaiaIPHAS_CMD}. One of these would have been entirely missed if proper motions were not taken into account (2MASSJ05493544$+$2329526 - Gaia DR2 3427482725113315200 - IPHAS DR2 J054936.18$+$232944.2), whilst the other would have been miss-matched using a 5'' cross-match between \gaia\ (epoch J2000) and IPHAS (2MASS J18592797$+$0156026 - Gaia DR2 4268571049773025024 - IPHAS DR2 J185928.18$+$015558.4). For the latter, we additionally mark the change in position within the CMD between the wrong and correct match. Figure \ref{fig:missmatchGaia} shows, for each of the above mentioned targets, the IPHAS $r$-band image. Also marked are the 1'' cross-match circles centred on the IPHAS recorded positions (dashed circles) and the 5'' circles centred on the \gaia\ J2000.0 coordinates (solid circles).

In the case of 2MASSJ05493544$+$2329526, no match is found when rewinding the \gaia\ astrometry to epoch J2000 because, when the IPHAS observation was made, the target had already left the 5'' radius. However, taking the IPHAS epoch of observation into account, together with the \gaia\ proper motions, reveals 2MASSJ05493544$+$2329526 to lie on the cold end of the white dwarf track at a distance of just under 40 parsec and moving with a transverse velocity of just under 100 km/s.

In another case, 2MASSJ05210188$+$3425116 had also left the 5'' \gaia\ target radius at IPHAS epoch of observation. However, because another faint source happened to lie within that same radius, a different IPHAS source had been associated to the \gaia\ target. Correcting for the IPHAS observation epoch reveals 2MASS J18592797$+$0156026 to lie on the M-dwarf end of the main sequence, at a distance of $\approx 25$ parsec, and to be moving with a transverse velocity of $\approx 80$ km/s. 

The value-added catalogues contain many more correctly recovered targets which can now be inspected with high confidence similarly to the above mentioned examples.

\subsection{Identifying ``hidden'' H$\alpha$ excess sources}

\begin{figure*}
	\includegraphics[width=1\columnwidth]{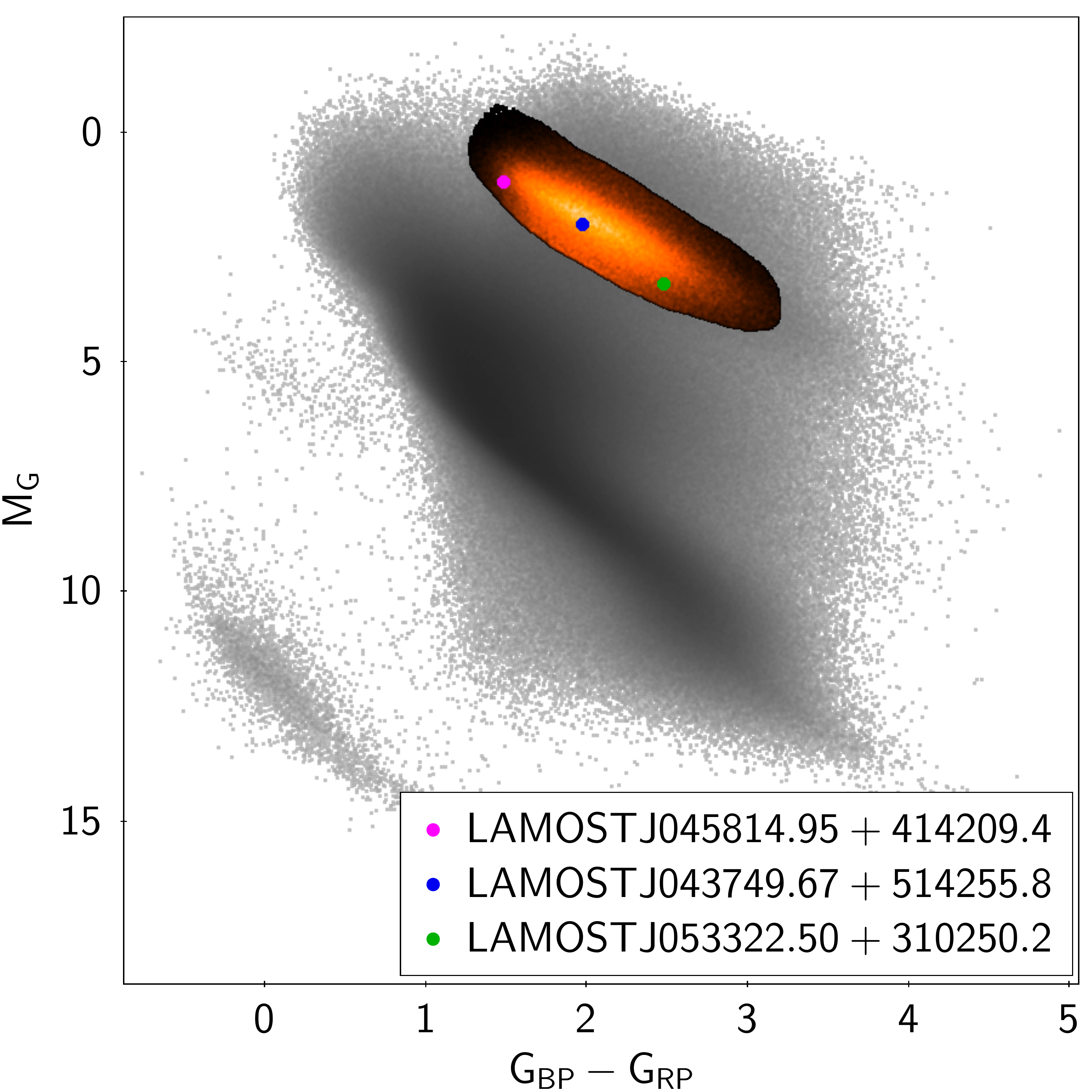}
    \includegraphics[width=1\columnwidth]{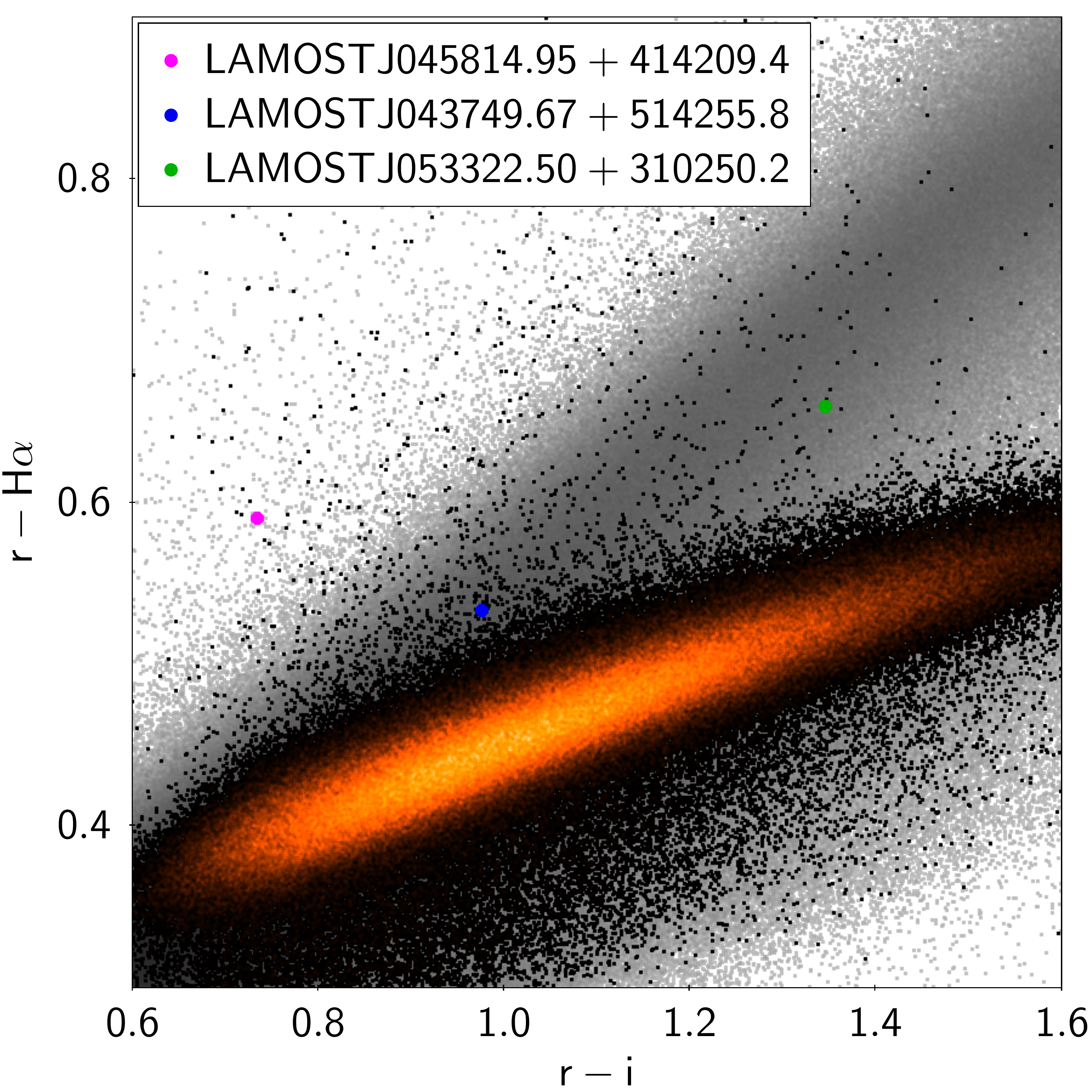}
    \caption{\gaia-based CMD and IPHAS colour-colour diagram for targets in our \gaia/IPHAS catalogue after applying our recommended quality cuts (grey points). Highlighted in red/black is the position of the reddened red-clump track within the \gaia\ CMD and its corresponding location in the IPHAS colour-colour diagram. Additionally marked are the locations of 3 objects which we discuss in the text, and for which LAMOST spectra exist (see Figure \ref{fig:LAMOST_spectra}).}
    \label{fig:LAMOST_CMD}
\end{figure*}

One of the biggest strengths of IPHAS (as well as KIS) is the inclusion of the narrow H$\alpha$ photometric band.  This has two uses: (i) the $(r - H\alpha)$ colour, combined with a  broadband colour such as $(r - i)$ or $(g - i)$, enables a fix on intrinsic colour and extinction for the majority of detected sources \citep[see the discussion of this and its exploitation for extinction mapping initiated by][]{sale09}, (ii) large numbers of emission line stars are made evident when the $(r - H\alpha)$ colour is strong enough to represent an excess relative to the main stellar locus -- removing the need for large-scale spectroscopic surveys (see e.g. \citealt{witham06,witham07,witham08,raddi13,scaringi13,gkouvelis16}). However, prior to the release of \gaia\ DR2, the selection of H$\alpha$-excess source candidates has usually been relative to the population of targets within a specific patch of sky, making no distinction between different luminosity classes. Most previously selected H$\alpha$-excess sources have been identified because their H$\alpha$ excess places them above the unreddened main-sequence on the IPHAS $r-i$ vs $r-$H$\alpha$ colour-colour diagram. An obvious drawback is that distant H$\alpha$-excess systems -- especially those behind large extinction columns -- will not stand out at moderate or small H$\alpha$-excess, since such objects will have the same colours as less reddened, later-type main-sequence stars in the ($r$-$i$) vs. ($r$-H$\alpha$) colour plane (see Fig. \ref{fig:LAMOST_CMD}, right panel).

The additional parallax (and thus distance) information provided by \gaia\ DR2 now provides luminosity information that allows us to select H$\alpha$-excess sources relative to specific regions in CMD space. To illustrate the potential of this method, we have manually selected objects lying on the red-clump reddening strip from our cleaned value-added catalogue in Figure \ref{fig:LAMOST_CMD}, left panel. The corresponding IPHAS colours are plotted in the right panel of this figure. 

A number of sources within this subset of systems clearly exhibit H$\alpha$-excess. Although a proportion of these H$\alpha$-excess sources would have been picked out previously as they lie above the unreddened main sequence, many others are ``hidden'' below it. Using the \gaia\ distance information, and selecting H$\alpha$-excess sources based on a selection of higher-luminosity targets from the \gaia\ CMD, allows the identification of these systems as H$\alpha$-excess sources. 

We illustrate this using 3 examples marked in Figure \ref{fig:LAMOST_CMD} as solid circles along the reddened red-clump. These 3 systems have been selected since they are bright and have reliable LAMOST spectra (\citealt{lamost}), with a signal-to-noise greater that 10 in the r-band wavelength range and IPHAS photometry fainter than 13 in all bands (which ensures we are fainter than the non-linearity close to saturation). One of these is a clear H$\alpha$ emitter as it clearly lies above the unreddened main sequence track. The other two targets would have been difficult to identify as H$\alpha$-excess sources without the additional parallax/distance information since they fall below the unreddened main sequence track in the IPHAS colour plane. 
 
\begin{figure*}
	\includegraphics[width=2\columnwidth]{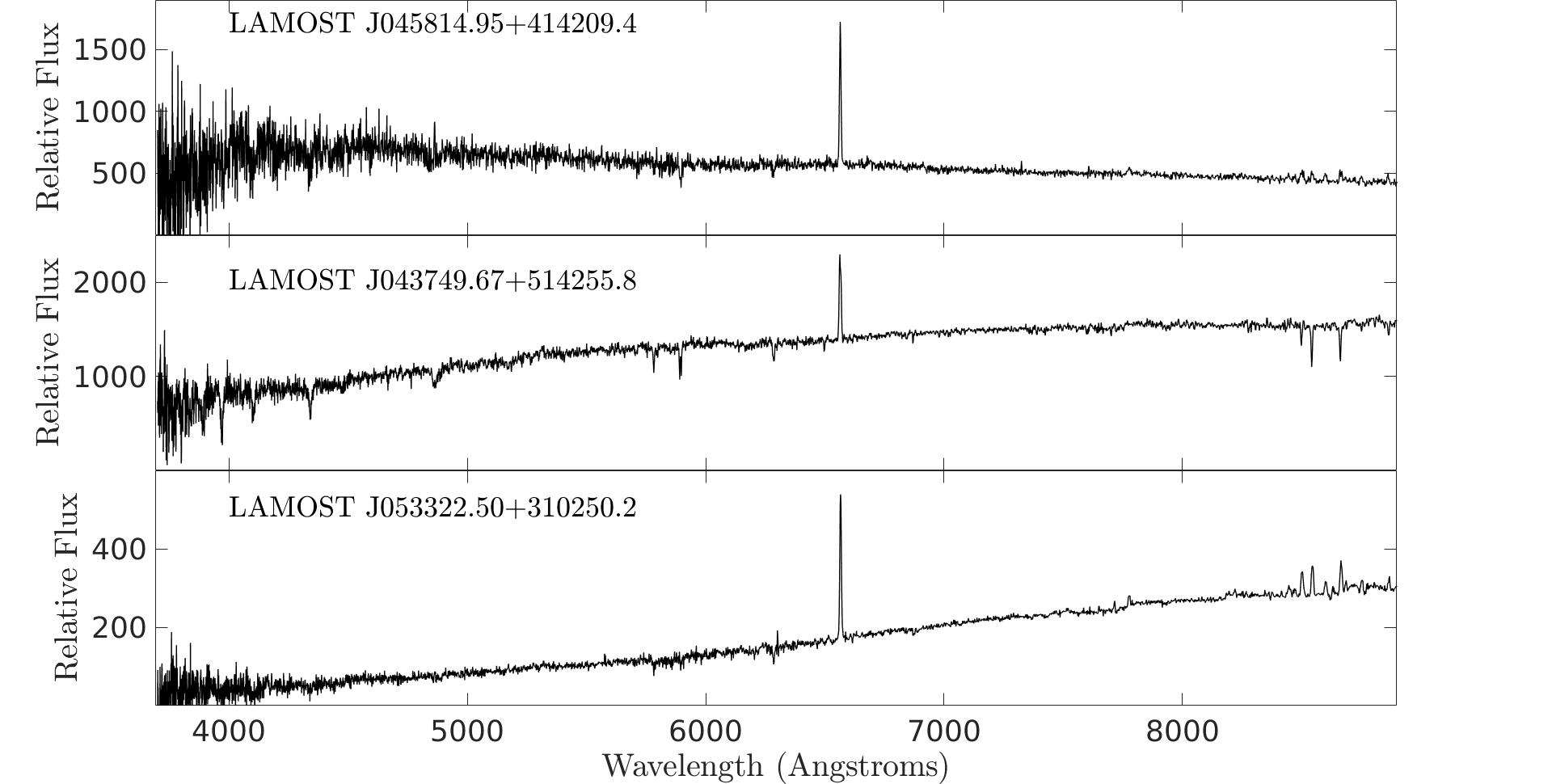}
    \caption{LAMOST spectra for the 3 objects picked out in Figure \ref{fig:LAMOST_CMD}.}
    \label{fig:LAMOST_spectra}
\end{figure*}

The LAMOST spectra for these 3 objects are shown in Figure \ref{fig:LAMOST_spectra}, and all present clear H$\alpha$ emission lines. Based on their spectra we classify both LAMOST J045814.95$+$414209.4 and LAMOST J053322.50$+$310250.2 as Be or Herbig stars with disks. LAMOST J043749.67$+$514255.8 is most probably an Ae star, given the combination of H$\alpha$ emission and strong Calcium triplet absorption in its spectrum. All three are rare objects in the catalogue in that they are early type objects with distances in excess of $\sim$2 kpc (based on their parallaxes), and visual extinctions, $A_V > 3$ (based on their IPHAS $(r - i)$ colours, assuming intrinsic colours of $\sim$0 or less).  

\subsection{Identifying new variable sources}
Our \gaia/IPHAS and \gaia/KIS value-added catalogues can also be used to identify new variable stars. 

As an illustration, we have inspected the $G_{RP}-i$ colour distribution of sources in our catalogues and selected targets with $|G_{RP}-i|>1$ . This provides a useful proxy for large amplitude variability, since the IPHAS $i$-band filter lies entirely within the Gaia $G_{RP}$-band one. There are 104 sources which satisfy both this variability criteria and the quality control cuts discussed in Section \ref{sec:clean}. 

As discussed in \cite{arenou18} and \cite{riello18}, the \gaia\ $G_{BP}$ and $G_{RP}$ fluxes are obtained from simple integration of a 3.5 by 2.1 arcsecond window, and it is possible that close sources contaminate this measurement. The \gaia\ team thus also provide a so-called \texttt{phot\_bp\_rp\_excess\_factor} value for each source, which tries to measure the excess flux of the $G_{BP}$ and $G_{RP}$ bands when compared to the broader $G$ band. Only one target out of the 104 sources found to be variable has a \texttt{phot\_bp\_rp\_excess\_factor}>2, with a sample mean of $1.45$. For comparison, a mean value of 1.35 is found for our whole catalogue satisfying our quality control cuts. Visual inspection of the fields for many of these candidate variable targets reveal that these are mostly isolated sources, but there are a few exceptions. These targets will have to be followed-up to be confirmed as true variables.

\section{The Gaia/IPHAS and Gaia/KIS Value-added Catalogues} \label{sec:VAC}
We provide our full value added catalogues without any quality selection cuts so that users can devise their own preferred selection based on $f_c$ and $f_{FP}$ if they so wish. These include 7,927,224 and 791,071 entries for \gaia/IPHAS \gaia/and KIS, respectively. 

In addition, we provide a ``light version'' of the value-added catalogues, with a reduced number of columns and our recommended general-purpose quality cuts already applied (see Section \ref{sec:clean}). These include 7,479,991 ($\approx 94\%$) and 773,464 ($\approx 98\%$) entries for \gaia/IPHAS \gaia/and KIS respectively. 

The column definitions for our value-added catalogue are provided in Appendix \ref{app:cols}. The same table notes whether the columns can be found in the \gaia/IPHAS or \gaia/KIS catalogues, as well as whether the column is included in our value-added light versions of the catalogues. All catalogues can be accessed online from the VizieR service.

\section{Conclusion} \label{sec:conc}
We have presented a sub-arcsecond cross-match between \gaia\ DR2 and both IPHAS and KIS. This was achieved by taking into account both the proper motions reported in \gaia\ as well as the observation epochs reported in both IPHAS and KIS. Our value-added catalogues also contain two additional quality control parameters: a so-called ``completeness'' fraction ($f_c$) which provides information relating to how acceptable the \gaia\ astrometric solution is compared to targets with similar $G$-band magnitudes, and a so-called false-positive fraction ($f_{FP}$) providing information on how reliable the astrometric measurements of a given target are. We provide both the full catalogues with all entries, as well as a light version containing targets which satisfy our preferred quality control cuts. Aside from providing the additional IPHAS and KIS photometry to the \gaia\ information for individual targets, our catalogues can be used identify and select H$\alpha$-excess emission line sources in any part of the \gaia\ CMD as well as new variable targets. 

The full \gaia/IPHAS catalogue contains 7,927,224 targets of which 7,479,991 pass our quality control cuts, whilst the \gaia/KIS catalogue contains 791,071 targets of which 773,464 pass the quality cuts. Both the full and light versions of the catalogue can be obtained through VizieR.

\section*{Acknowledgements}
Cross-matching between catalogues has been performed using STILTS, and diagrams were produced using the astronomy-oriented data handling and visualisation software TOPCAT (\citealt{topcat}). This work has made use of the Astronomy \& Astrophysics package for Matlab (\citealt{matlabOfek}). This research has also made extensive use of the SIMBAD database, operated at CDS, Strasbourg. JED and MM acknowledge the support of a research grant funded by the Science, Technology and Facilities Council of the UK (STFC, ref. ST/M001008/1).




\bibliographystyle{mnras}
\bibliography{gaiaCat} 


\newpage
\appendix

\onecolumn
\section{Catalogue format} \label{app:cols}
\small
\begin{longtable}{p{1.5cm}p{1.5cm}p{1.5cm}p{1.5cm}llp{5cm}}

\caption{\label{tab:cols} 
Definition of columns in the \gaia/IPHAS and \gaia/KIS catalogues. Column indices are given for both the light and full catalogue versions. Light versions only contain a limited number of columns and are restricted to sources satisfying our recommended quality cuts of $f_c$<0.98 and $f_{FP} \leq 0.02$. Absolute magnitudes and transverse velocities have been computed using the 1/parallax method. It is important to point out that the absolute magnitude and colour columns in the catalogue do not take extinction/reddening into account.}\\

\hline
\gaia/IPHAS (full) & \gaia/IPHAS (light)& \gaia/KIS (full) & \gaia/KIS (light) & Column name & Unit & Description \\
\hline
\endfirsthead

\multicolumn{3}{c}%
{{\bfseries \tablename\ \thetable{} -- continued}} \\
\hline
\gaia/IPHAS (full) & \gaia/IPHAS (light)& \gaia/KIS (full) & \gaia/KIS (light) & Column name & Unit & Description \\ 
\hline
\endhead

\hline \hline
\endlastfoot

1 & 1 & 1 & 1 & GaiaDR2& & Unique \textit{Gaia} DR2 source designation \\

2 & 2 & 2 & 2 & ra & degrees & \textit{Gaia} DR2 barycentric right ascension (ICRS) at Epoch 2015.5 \\

3 & 3 & 3 & 3 & dec & degrees & \textit{Gaia} DR2 barycentric declination (ICRS) at Epoch 2015.5 \\

4 & - & 4 & - & e\_ra & mas & Standard error of right ascension (raErr$\times$cos(dec)) \\

5 & - & 5 & - & e\_dec & mas & Standard error of declination \\

6 & 4 & 6 & 4 & Plx & mas & Absolute stellar parallax\\

7 & - & 7 & - & e\_Plx & mas & Standard error of parallax\\

8 & 5 & 8 & 5 & pmra & mas/yr & Proper motion in right ascension direction (raPM$\times$cos(dec)) \\

9 & 6 & 9 & 6 & pmdec & mas/yr &  Proper motion in declination direction\\

10 & - & 10 & - & e\_pmra & mas/yr & Standard error of proper motion in right ascension direction\\

11 & - & 11 & - & e\_pmdec & mas/yr & Standard error of proper motion in declination direction\\

12 & 7 & 12 & 7 & G & mag & \textit{Gaia} $G$-band magnitude\\

13 & 8 & 13 & 8 & BP & mag & \textit{Gaia} $G_{BP}$-band magnitude\\

14 & 9 & 14 & 9 & RP & mag & \textit{Gaia} $G_{RP}$-band magnitude\\

-  & - & 15 & 10 & U & mag & KIS $U$-band magnitude\\

-  & - & 16 & - & e\_U & mag & Standard error on KIS $U$-band magnitude\\

-  & - & 17 & 11 & g & mag & KIS $g'$-band magnitude\\

-  & - & 18 & - & e\_g & mag & Standard error on KIS $g'$-band magnitude\\

15 & 10 & 19 & 12 & r & mag & IPHAS or KIS $r'$-band magnitude\\

16 & - & 20 & - & e\_r & mag & Standard error on IPHAS or KIS $r'$-band magnitude\\

17 & 11 & 21 & 13 & i & mag & IPHAS or KIS $i'$-band magnitude\\

18 & -  & 22 & - & e\_i & mag & Standard error on IPHAS or KIS $i'$-band magnitude\\

19 & 12 & 23 & 14 & ha & mag & IPHAS or KIS H$\alpha$-band magnitude\\

20 & - & 24 & - & e\_ha & mag & Standard error on IPHAS or KIS H$\alpha$-band magnitude\\

21 & - & 25 & - & BPmRP & mag & $G_{BP}-G_{RP}$ colour\\

22 & - & 26 & - & rmi & mag & $(r' - i')$ colour\\

23 & - & 27 & - & rmha & mag & $(r' - $H$\alpha)$ colour\\

-  & -  & 28 & - & Umg & mag & $(U - g')$ colour\\


24 & - & 29 & - & pmT & mas/yr & Transverse proper motion\\

25 & - & 30 & - & vT & km/s & Transverse velocity\\

26 & 13  & 31 & 15 & mMJD & d & Modified Julian Date used for cross-matching \textit{Gaia} to IPHAS and KIS.\\

27 & -  & 32 & - & mMJD\_separation & arcsec & Angular separation between the rewinded \textit{Gaia} position at Epoch mMJD to the nominal IPHAS or KIS position\\

28 & - & 33 & - & M\_G & mag & Absolute \textit{Gaia} $G$-band magnitude (not corrected for extinction)\\

29 & - & 34 & - & M\_r & mag & Absolute IPHAS $r'$-band magnitude (not corrected for extinction)\\

-  & -  & 35 & - & M\_U & mag & Absolute KIS $U$-band magnitude (not corrected for extinction)\\

30 & -  & 36 & - & redChi2 &  & Reduced $\chi^2$ ($\chi_{\nu}^2$) obtained from the \textit{Gaia} astrometric fit\\

31 & 14 & 37 & 16 & f\_c &  & Retention fraction based on $\chi_{\nu}^2$ (completeness)\\

32 & 15 & 38 & 17 & f\_FP &  & False-Positive rate based on negative parallax sample\\

33 & -  & 39 & - & ra\_mMJD & degrees & Right ascension provided by IPHAS or KIS DR2 (ICRS, Epoch mMJD)\\

34 &  - & 40 & - & dec\_mMJD & degrees & Declination provided by IPHAS or KIS DR2 (ICRS, Epoch mMJD)\\

35 &  - & -  & - & fieldID &  & IPHAS DR2 field identifier\\

36 &  - & - & - & IPHAS\_name &  & IPHAS DR2 name\\

- &  - & 41 & - & KIS\_name &  & KIS DR2 name\\

- &  - & 42  &- & KIC &  & KIC ID \\

\end{longtable}
\normalsize
\twocolumn


\bsp	
\label{lastpage}
\end{document}